\documentclass[aps,pra,twocolumn,superscriptaddress,floatfix,longbibliography]{revtex4-2}
\usepackage{amsmath}
\usepackage{amssymb}
\usepackage{bm}
\usepackage{graphicx}
\usepackage{color}
\usepackage[normalem]{ulem}
\usepackage{soul}
\usepackage{url}

\usepackage{verbatim}

\usepackage{hyperref}
\hypersetup{pdftex,colorlinks=true,allcolors=blue,bookmarksnumbered=true}
\usepackage{hypcap}

\usepackage{graphicx}
\usepackage{bm}
\usepackage{float}
\usepackage{booktabs}
\usepackage{tabularx}
\usepackage{colortbl}
\usepackage{etoolbox}
\usepackage{subfig}
\usepackage{soul}

\begin{document}

\title{Generation of Bimodal Solitons in a Sapphire Whispering Gallery Mode Maser at Millikelvin Temperatures}

\author{Catriona A. Thomson}
\affiliation{ARC Centre of Excellence for Engineered Quantum Systems and ARC Centre of Excellence for Dark Matter Particle Physics, Department of Physics, University of Western Australia, 35 Stirling Highway, Crawley WA 6009, Australia}

\author{Michael E. Tobar}
\affiliation{ARC Centre of Excellence for Engineered Quantum Systems and ARC Centre of Excellence for Dark Matter Particle Physics, Department of Physics, University of Western Australia, 35 Stirling Highway, Crawley WA 6009, Australia}

\author{Maxim Goryachev}
\email{maxim.goryachev@uwa.edu.au}
\affiliation{ARC Centre of Excellence for Engineered Quantum Systems and ARC Centre of Excellence for Dark Matter Particle Physics, Department of Physics, University of Western Australia, 35 Stirling Highway, Crawley WA 6009, Australia}

\date{\today}
\begin{abstract}
We present experimental observations of bimodal solitons in a solid state three-level maser cooled to millikelvin temperatures. The maser is built on a highly dilute $\textrm{Fe}^{3+}$ spin ensemble hosted by a high purity $\textrm{Al}_{2}\textrm{O}_{3}$ crystal constituting a high quality factor whispering-gallery-mode resonator. The maser is pumped through one of these modes, near 31 GHz, generating two signals near 12.04 GHz from two distinct modes, 8 MHz apart. The system demonstrates three regimes, namely, a continuous wave regime, a dense soliton regime and a sparse soliton regime. These results open new avenues for studying nonlinear wave phenomena using microwave systems as well as new applications of solitons in this part of the electromagnetic spectrum. 
\end{abstract}
\maketitle

\section{Introduction}
Optical frequency combs have become key in myriad applications from metrology and precision spectroscopy to tests of beyond the standard model physics, with their remarkable ability to compare clocks within 10$^{-19}$~Hz~\cite{Udem2002,Diddams2020}. Initially produced via mode-locked lasers~\cite{Kourogi1993,Udem1999,Reichert1999}, which revolutionized the field of frequency metrology~\cite{Diddams2007,Witte2005}, a particularly active area of current research is in developing alternative production mechanisms for such combs, in engineering systems balanced between nonlinear gain and dissipation.
\begin{figure}[t]
     \begin{center}
            \includegraphics[width=0.45\textwidth]{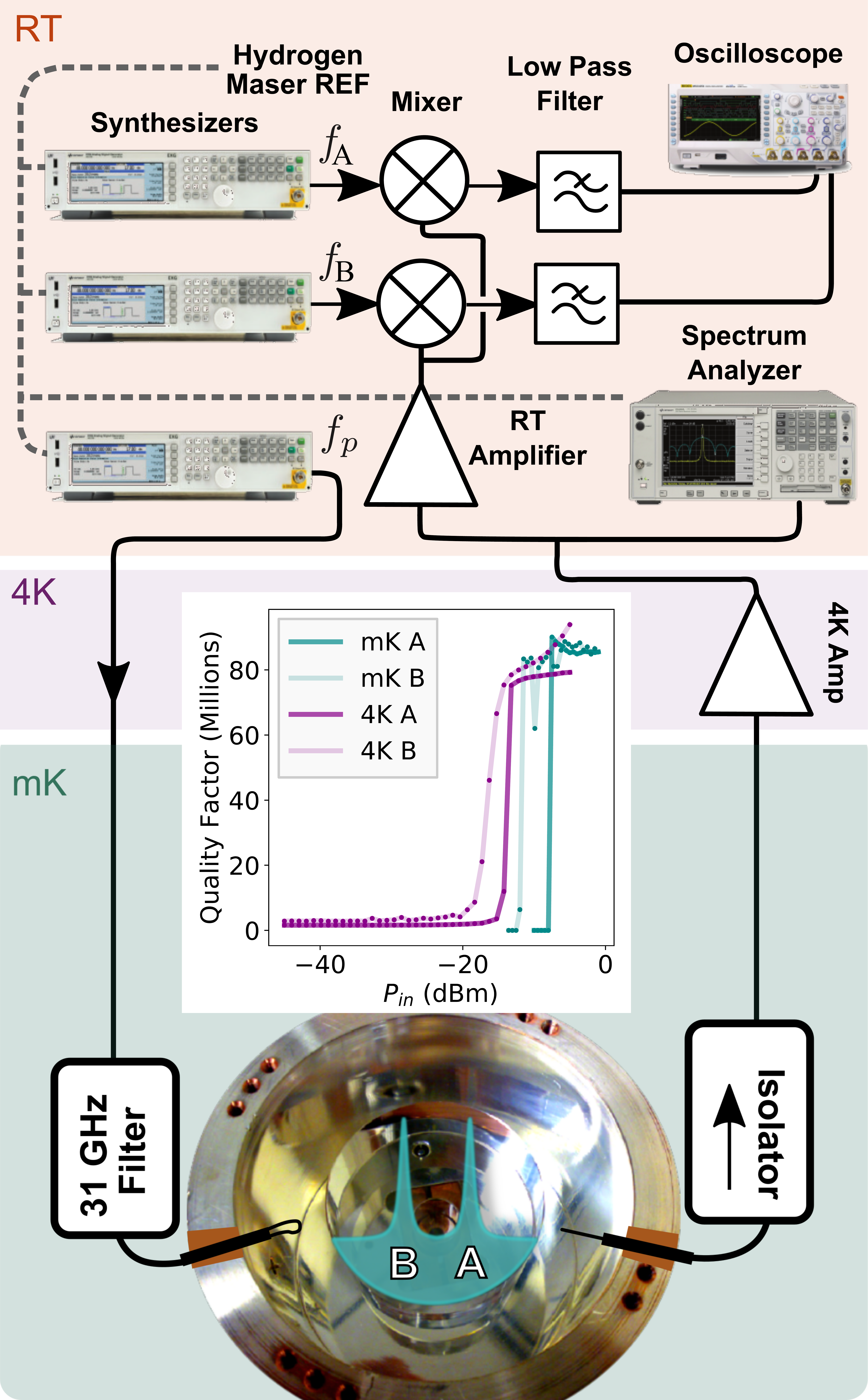}
            \end{center}
        \vspace{-5 mm}
    \caption{Experimental setup used to generate maser solitons at mK temperatures. A WGM crystal resonator is pumped from a room temperature signal source (around 31~GHz) through an input filter ensuring minimal excitation of the readout mode. Generated maser signals (around 12~GHz) are collected with a single antenna and fed through a low temperature isolator and a low noise cryogenic amplifier. These signals are analysed via a spectrum analyser and down converted via two mixers to be observed via an oscilloscope. Inset: Fitted quality factors of modes A and B at 4 K and mK, at a range of powers incident on the modes, without any applied pump mode. These whispering gallery modes require a certain threshold of incident power to be applied to the mode for high quality factor resonance to be possible.}
   \label{setup}
\end{figure}

A frequency comb may occur in conjunction with solitons within a resonator, equivalent to ultrashort phase-locked pulses circulating the cavity in the time domain. Such pulses may be infinitely supported within the resonator, given a continuous-wave pump source and a resonant medium which displays a balance between dispersion and non-linear gain \cite{Yi2015}. To this end, the nonlinear Kerr effect has been much exploited as the dissipative structure in various material systems~\cite{Yi2015,Barthelemy1985,Cambournac2002} such as sillica optical fibres~\cite{Leo2010}, and high-Q crystalline and silicon nitride-based micro-resonators \cite{Yi2015,Herr2013}.

Despite such plethora of systems demonstrating combs and solitons~\cite{Fortier:2019vc}, they are quite unique to the optical frequency domain. This article reports the first, to the best of authors' knowledge, observation of microwave solitons in a maser system. The reported phenomena was produced via the iron paramagnetic impurities (order of parts per billion) within the most pure HEMEX grade synthetic sapphire crystal, when cooled to millikelvin temperatures and with zero applied magnetic field \cite{Bourgeois2005}. The paramagnetic impurities have been shown to offer non-linear gain in form of a $\chi^3$ non-linearity at 4K in temperature, which besides masing \cite{Bourgeois2005}, has produced four wave mixing \cite{Creedon2012}, frequency conversion \cite{Creedon2012b} and time reversal symmetry breaking \cite{karim1,Goryachev:2014aa}. Dispersion arises from frequency dependence of the effective permeability and thus refractive index due to these paramagnetic impurities~\cite{Goryachev:2014aa} which can be altered by pumping a mode near the ion pump transition. Furthermore, at millikelvin temperature, effects of paramagnetic impurities on the properties of cooled sapphire resonators have been shown to be enhanced when compared with the effects at 4K as the majority of the spins condense to the ground state \cite{Goryachev:2014aa,Creedon:2011wk,PhysRevB.88.224426}. This work reports the first operation of such a maser at millikelvin temperatures, with the observation of new effects not seen in the 4K system, providing a new avenue for implementing optical metrological methods in the microwave and radio- frequency domains.

\section{Experimental Setup}
The experimental setup is based on a cylindrical HEMEX grade sapphire (undoped monocrystalline $\alpha$-Al$_2$O$_3$) resonator supporting extremely high quality factor whispering gallery modes (WGMs) in the microwave frequency range. The crystalline lattice of sapphire is host to a number of naturally occurring impurities including Fe$^{3+}$, Cr$^{3+}$, and Ti$^{3+}$ with typical concentrations of parts-per-billion in HEMEX-grade crystals~\cite{PhysRevB.88.224426}. Our crystal has been annealed to increase the Fe$^{3+}$ population by conversion of Fe$^{2+}$ to Fe$^{3+}$, with a 100-fold improvement achieved in the Fe$^{3+}$ to Fe$^{2+}$ ratio \cite{Creedon:2009aa}. Paramagnetic Fe$^{3+}$ ions in sapphire, each substituting an Al$^{3+}$ cation, was proposed ~\cite{KorPro} and implemented as a gain medium for cryogenic masers, which do not require any external magnetic field~\cite{Benmessai:2008aa,Creedon:2009aa} . Profitably, at zero applied field, this ion represents a three level system with the spin-$|1/2\rangle$ (ground state), $|3/2\rangle$ (intermediate state), and $|5/2\rangle$ (higher excited) states. These three levels are ideally suited to be used as a $\Lambda$-scheme: pumping the $|1/2\rangle\rightarrow|5/2\rangle$ transition via a high-frequency pump WGM within the ion's electron spin resonance (ESR) linewidth, inducing a fast non-radiative transition to the intermediate level with a short life time followed by a radiative maser transition $|3/2\rangle\rightarrow|1/2\rangle$, stimulating photon emission at the ESR frequencies around $\sim 12.04$ GHz, which happens to encompass the linewidth of two signal WGMs in the crystal (see Fig~\ref{setup}). These signal WGMs have sufficient quality factors to self-sustain resonance for sufficient pump power. Fig.~\ref{setup} (inset) shows the quality factors of modes A and B, displaying a very strong nonlinearity in the form of quality factor dependence on power incident on the mode.

The sapphire WGM crystal was housed in a silver-plated copper cavity and interrogated by two microwave probes inserted through the cavity walls. The output probe is an antenna oriented to couple to two modes in the region of $12.04$~GHz (specifically, we investigate the responses of mode A at 12.03812 GHz and mode B at 12.02979 GHz). The input probe is an orthogonally oriented loop which, ideally, strongly couples to the azimuthal magnetic component of one of the (dominantly transverse-magnetic) modes in the vicinity of the $31.9$~GHz transition and does not load quality factors of the signal modes, which are quasi-transverse-electric modes.

The cavity is attached to the millikelvin stage of a dilution refrigerator. The pump signal is generated by a microwave synthesizer locked to a hydrogen maser and delivered to the cavity via a series of copper and phosphor-bronze cables. The signal is cleaned by a narrow band-pass filter tuned to near $31.5$~GHz and directly bolted to the cavity. The total attenuation of the cables and the filter from the output of the synthesizer to the input of the cavity is $31$~dB at the pump frequency. The output signals from the crystal pass through a millikelvin isolator and a 4~K low noise amplifier. At room temperature the target signal is split between a spectrum analyzer, used to collect frequency data around the target mode, and two downconversion channels, used to mix down the signal for collection of time-series data. To achieve downconversion, two synthesizers generate frequencies of $f_{A}$ and $f_{B}$ to bring the signal carriers to the kHz level whereby they are individually digitized with a two-channel high-speed real-time oscilloscope. A schematic of the experimental setup used to observe maser solitons is shown in Fig.~\ref{setup}.

\section{Observations}
\begin{figure}
     \begin{center}
            \includegraphics[width=0.5\textwidth]{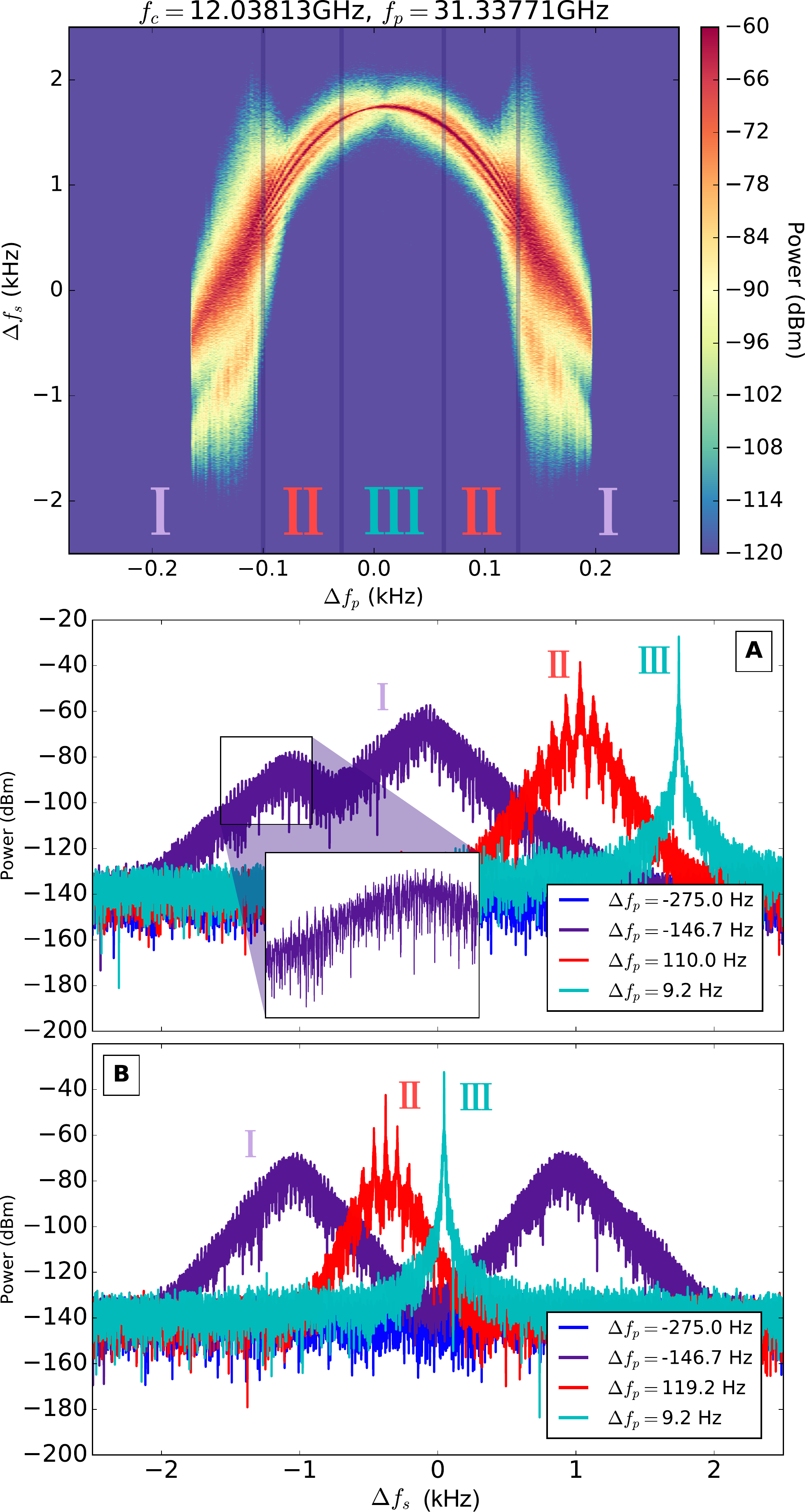}
            \end{center}
        \vspace{-5 mm}
    \caption{Top: Spectra of masing on mode A ($f_s$) for different pump frequencies around $31.3377$~GHz ($\Delta f_p$). A/B: snapshots of masing on mode A/B at various $\Delta f_p$, illustrating the three masing regimes. Note that $f_s$ is arbitrarily chosen, serving only to indicate scale, as the signal frequency has no unambiguously natural centre.}
   \label{mKColour}
           \vspace{-5 mm}
\end{figure}
 \begin{figure}
     \begin{center}
            \includegraphics[width=0.5\textwidth]{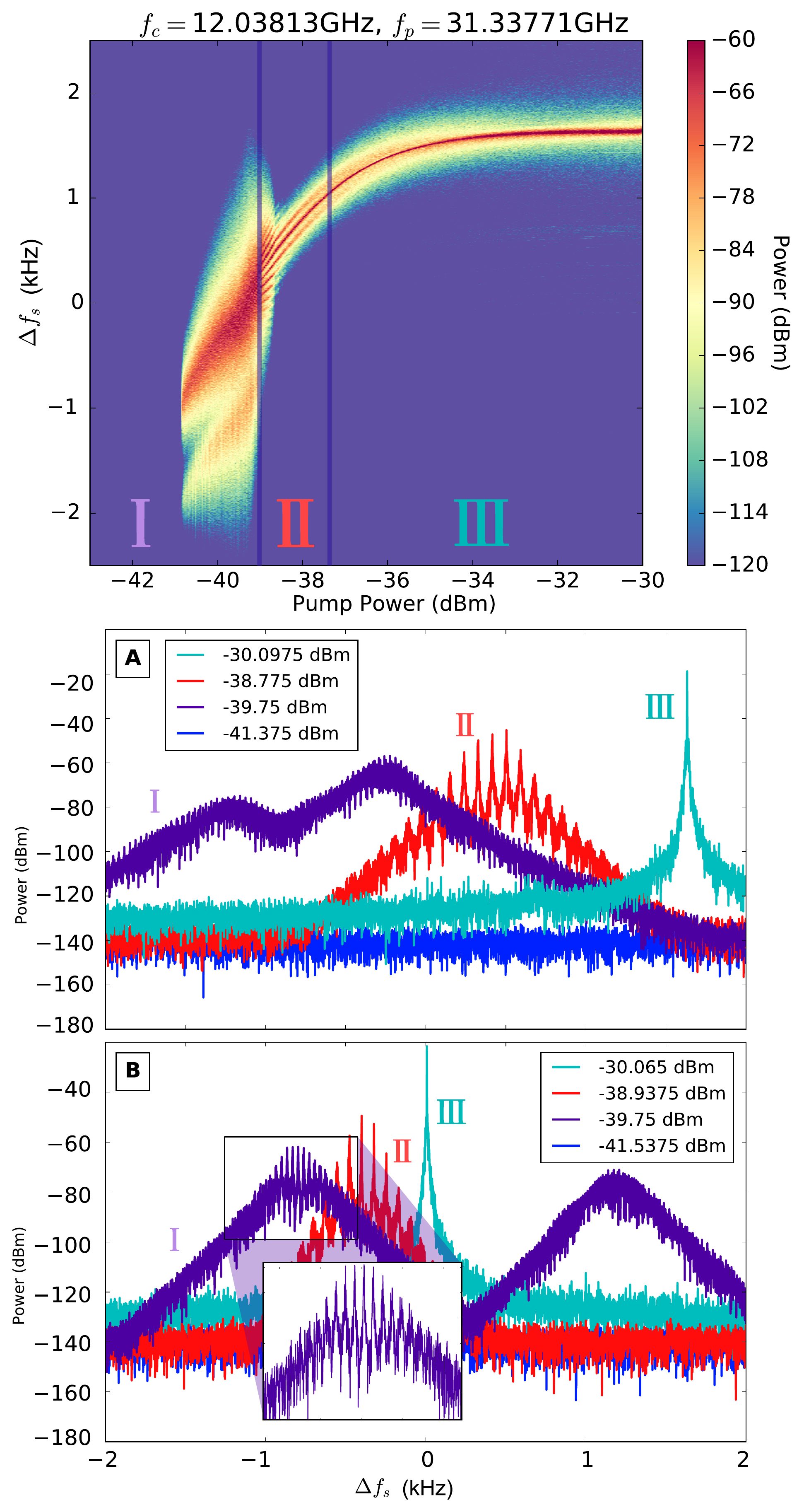}
            \end{center}
        \vspace{-5 mm}
    \caption{Top: Spectra of masing on mode A ($f_s$) for different pump powers at $31.3377$~GHz. A/B: snapshots of masing on mode A/B at various pump powers, illustrating the three masing regimes. Note that $f_s$ is arbitrarily chosen, serving only to indicate scale, as the signal frequency has no unambiguously natural centre.}
   \label{mKPColour}
           \vspace{-5 mm}
\end{figure}
Masing was observed on two signal modes ($f_A = 12.03813$~GHz and $f_B = 12.02979$~GHz) for a range of pump frequencies, $\Delta f_p$, swept around the pump mode, $f_p = 31.33771$~GHz. Full spectra of the masing signal from $f_A$ is given in Fig~\ref{mKColour}, along with snapshots from both $f_A$ and $f_B$.
\begin{table}[hbt] 
\centering
\captionsetup{font={bf},skip=0.5ex}
\caption*{Target Whispering Gallery Modes}
\begin{tabularx}{0.5\textwidth}{X X X X}
\toprule
 \bfseries Pump Mode  & \bfseries Frequency (GHz) &  \bfseries Readout Mode  & \bfseries Frequency (GHz)\\
\midrule
1     & 31.33771     & A & 12.03812\\
2   & 31.33974 &    B     & 12.02979\\ \bottomrule
\end{tabularx}
\end{table}
 \begin{figure*}
     \begin{center}
            \includegraphics[width=1\textwidth]{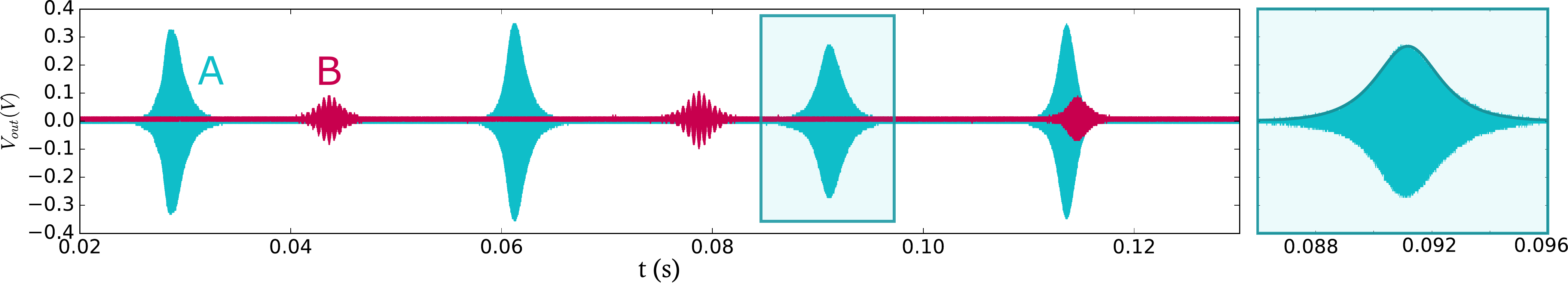}
            \end{center}
        \vspace{-5 mm}
    \caption{Time-series output of mode A (blue) and B (pink) of the crystal, in regime I, being pumped at $31.337707136$~GHz, after being mixed down to baseband, representing modulation of the carrier mode. The soliton pulse train is the time-series equivalent of the comb structure. The fitted \textit{sech} envelope of one pulse is highlighted.}
   \label{solitons}
           \vspace{-5 mm}
\end{figure*}
 \begin{figure}
     \begin{center}
            \includegraphics[width=0.48\textwidth]{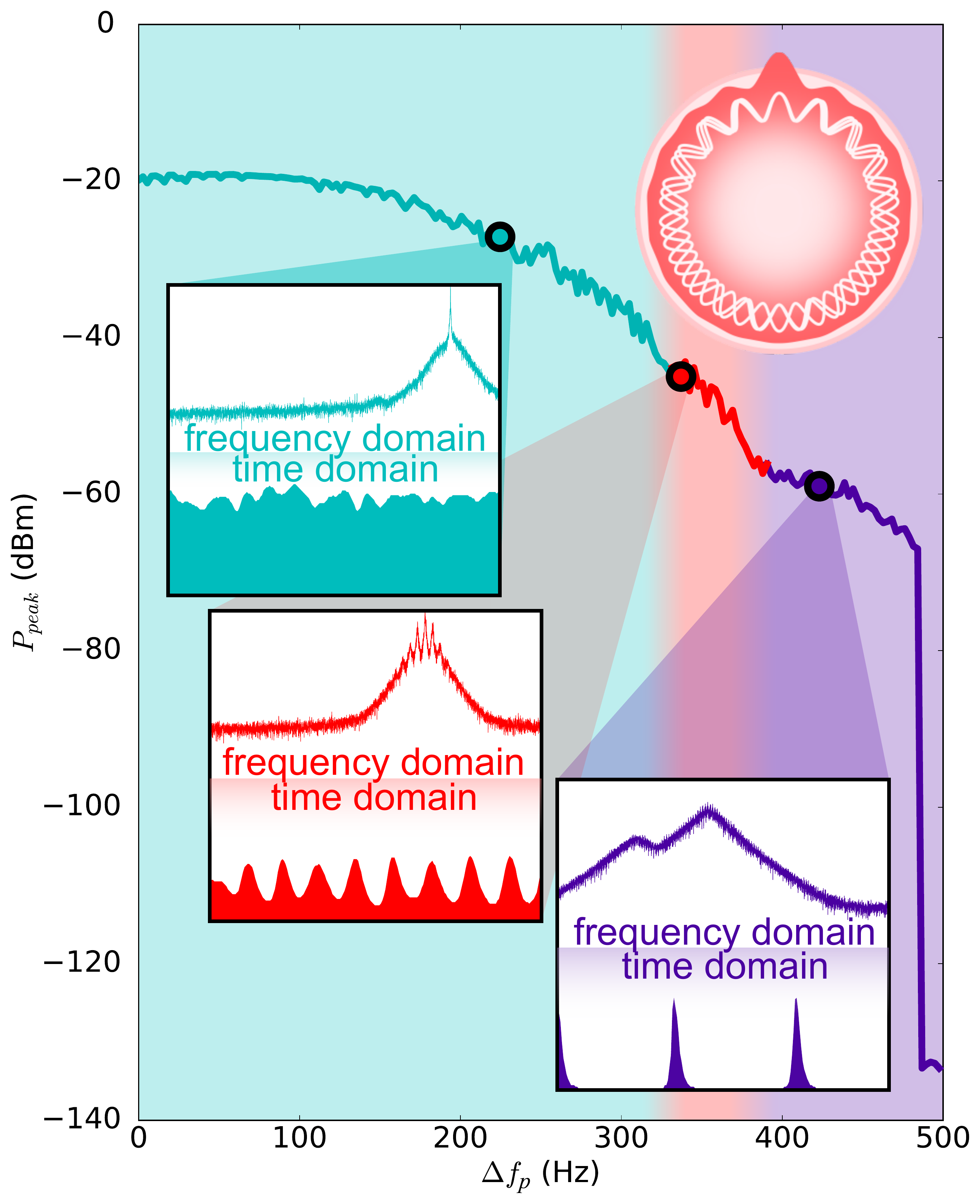}
            \end{center}
        \vspace{-5 mm}
    \caption{Maser peak emission ($P_{peak}$) for various pump frequencies about $31.3377$~GHz ($\Delta f_{p}$) for mode A, with $P_{pump} = 0$~dBm. $P_{peak}$ is defined as the amplitude of the centre of the dominant peak in the maser spectrum. Three points are highlighted, one for each regime. The inset panels show maser spectra and associated soliton outputs in the various regimes;  (III) single spectral line regime (blue), (II) the dense soliton regime (red), and (I) the sparse soliton regime (purple). Discrete jumps between the regimes can be observed on the spectrum of peak maser output. Top-right inset illustrates the relationship between the comb frequencies, and the intracavity soliton whose existence implies synchronised phases between the comb frequencies.}
   \label{solitonmap}
           \vspace{-5 mm}
\end{figure}

\subsection{Comb Structure}
At millikelvin temperatures (within a dilution fridge cooled to $\sim$16~mK, but transiently heated to $\sim$20-50~mK), a comb-like structure was obtained, the first such structure observed being derived from a maser system to the knowledge of the authors (Figs.~\ref{mKColour} and \ref{mKPColour}, inset). Comb lines only persist down to approximately -40 dBm incident pump power, below which the comb collapses. Data at 4~K were also obtained, but results are deferred to the appendices. It has to be noted that the 4K and mK regimes for the maser system are sufficiently different. The occupation number of the excited state for the system at 4K is ${n}= \frac{1}{e^{\frac{h\nu}{k_B T}}+1}\approx 0.41$ whereas for 50mK it is $8.643\times10^{-14}$.

Fig.~\ref{mKColour} and \ref{mKPColour} demonstrate three distinct generation regimes: (III) the continuous wave regime at the centre (near zero $\Delta f_p$) with a narrow masing peak, (II) the few peak regime for larger detuning, corresponding to high density soliton emission, followed by (I), a sparse soliton regime just surpassing the masing threshold on both sides of the masing regime, with few solitons. The latter regime demonstrates a second broad feature that can be attributed to the doublet phenomenon in WGM systems where the solution has two degenerate components.  Various symmetry breaking features in the sapphire resonator lift this degeneracy and our readout mode thus exists as a double peaked structure. It is worth noting that for the continuous wave regime with $\Delta f_p\sim 0$, one peak supercedes the other within the doublet. The dependence of the masing regime upon explicitly pump power, at a fixed pump frequency, was also clearly observed (see Fig.~\ref{mKPColour}), displaying analogous regimes to the detuned frequency experiment. Due to the lineshape of the WGM pump mode, pumping off resonance delivers less power to the pump transition; hence, either parameter, pump power or pump frequency, can be used to control power delivered to the pump transition. It should be noted that the ``bimodality'' of the system refers, in fact, to the simultaneous masing on modes A and B, which are separated by $\sim$ 8 MHz, and not to the doublet phenomenon caused by lifted degeneracy.

\subsection{Pulse Train}
Examining the time-series data, this masing can be observed to be expelled in discrete millisecond-scale solitons, the time-domain phenomenon associated with the observed comb structure (see Fig.~\ref{solitons}). The emission of discernible pulses can be observed up to a saturation point, corresponding to single-frequency maser emission (cyan in Fig.~\ref{mKPColour}). The relationship between comb shape and soliton emission period is illustrated in Fig.~\ref{solitonmap}, which maps out the three regimes, depeneant upon pump frequency/power.

\section{Theoretical Description of the Maser Soliton System}
\begin{figure}[h!]
\includegraphics[scale=0.4]{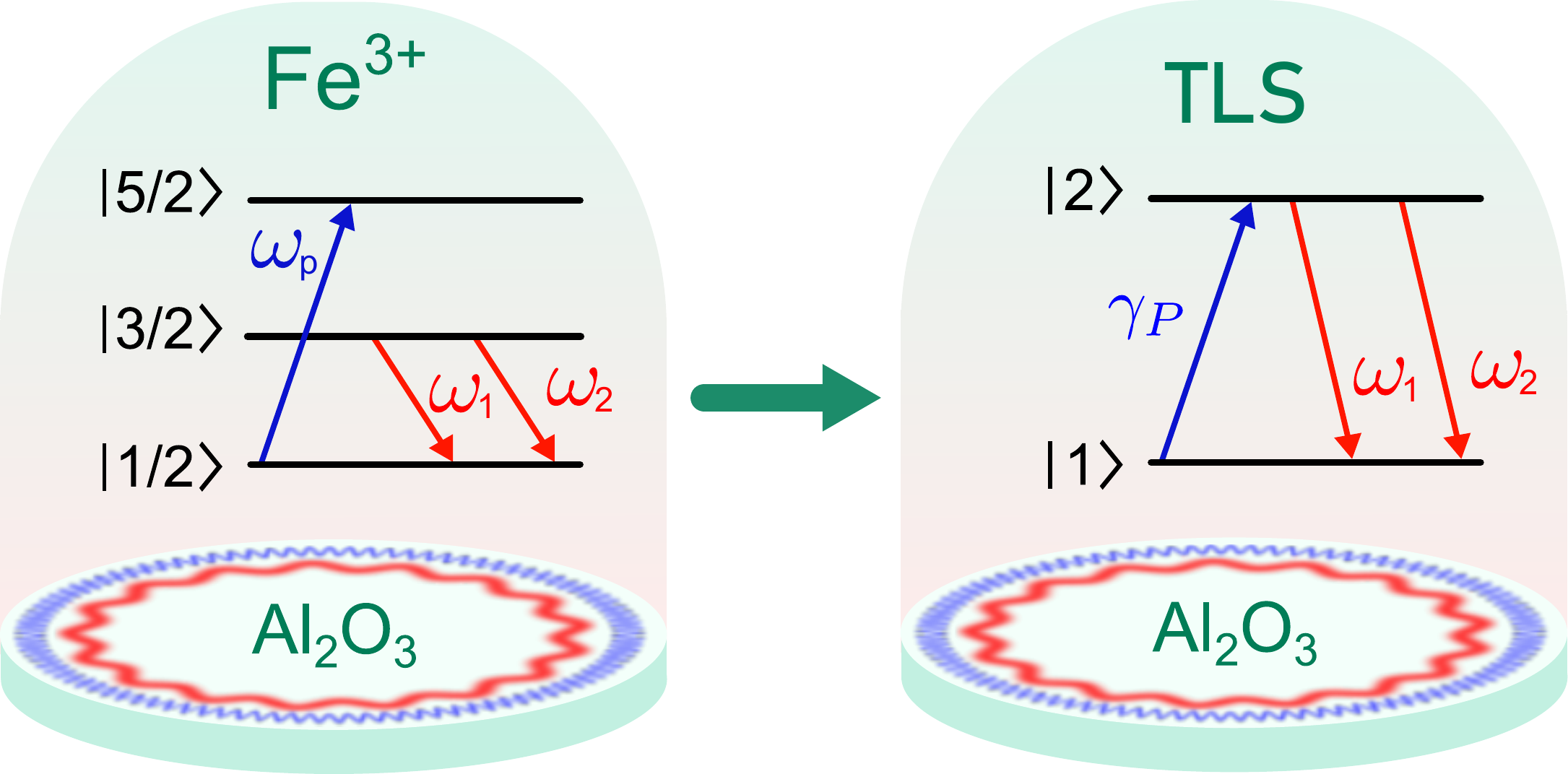}
\caption{Schematic of the three-level spin system of the Fe$^{3+}$ ion (left) which leads to masing. Right shows a simplification of the system to a two level system which is justified by theoretical and experimental observations.}
\label{TLS}
\end{figure}

\vspace{3mm}
In a host lattice of sapphire, Fe$^{3+}$ ions effectively act as a three level system with $|1/2\rangle$ being the ground state, $|5/2\rangle$ the excited state and $|3/2\rangle$ the intermediate state. A diagram of this ion is shown in Fig.~\ref{TLS}. The maser works by pumping ions from the ground state to the exited state from where they then decay into the intermediate state. The stimulated emission between $|3/2\rangle$ and $|1/2\rangle$ leads to masing.
\vspace{3mm}

The full description of the system discussed in this work includes a cavity pump mode (whispering gallery modes (WGM) at 31.33771 GHz or 31.33974 GHz), two signal modes (WGMs at 12.03812 GHz and 12.02979 GHz), an ensemble of three level systems, and associated dissipation baths. Generation of solitons in such $\Lambda$ systems have been considered before in the context of electromagnetically induced transparency (EIT) \cite{Wu2004}. In this work, the system is different to EIT as it represents a $\Lambda$ scheme maser.
\subsection{Simplified System Hamiltonian}
For the Fe$^{3+}$ ions in sapphire, the $|5/2\rangle$ state rapidly decays into the $|3/2\rangle$ state and the lifetime of the latter is considerably longer. So, observationally, the $N_{|5/2\rangle}$ population is near zero with total population $N \approx N_{|1/2\rangle} + N_{|3/2\rangle}$ at any given time. Furthermore, the dynamics of the pump mode can be adiabatically eliminated as it is dominated by the spin ensemble. Both signal modes are very high $Q$ relative to losses in the Fe$^{3+}$ ensemble, therefore the latter is the dominant source of losses. Additionally, we note that the signal modes are sufficiently far away from each other in frequency space, so that their direct coupling can be neglected. These facts lead to the significantly simplified description of the system as a dual mode laser system with an externally pumped TLS, with the crystal and the ensemble providing dispersion. The corresponding Hamiltonian (in units $\hbar=1$) is
\begin{equation}
	\label{MBE}
	 \begin{split}
 H=\omega_{a} \sum_{n} \sigma^{+}_n \sigma^{-}_n+\sum_{m=1}^2&\omega_{m} a_m^{\dagger} a_m\\+
  &\sum_{n,m}g_m\left(\sigma^{-}_n a^{\dagger}_m+\sigma^{+}_n a_m\right)
\end{split}
\end{equation}
where $m$ is a summation index over the photonic signal modes with resonance frequencies $\omega_m$ and creation (annihilation) operators $a^\dagger_m$ ($a_m$); $n$ is an index over the spin ensemble; $\sigma^{+}_n$, $\sigma^{-}_n$ and $\sigma^{z}_n$ are the usual Pauli matrices and $g_m$ are the spin-photon coupling constants. The system is completed by introducing photonic mode loss rates $\gamma_m$ and TLS emission, decoherence and pump rates: $\gamma_E$, $\gamma_D$ and $\gamma_P$ respectively. 
\subsection{Relationship to the Lugiato-Lefever Equation}
In the limit of a large TLS ensemble, the corresponding Maxwell-Bloch equations (see Chap. 11 in \cite{Lugiato2015}) could be written as follows:
\begin{equation}
\begin{aligned}
	\label{MBE}
	 \begin{array}{c}
\displaystyle \frac{\partial F_m}{\partial t} + \widetilde{c}\frac{\partial F_m}{\partial z}=\left(i \omega_m-\gamma_{m} / 2\right) F_m-i g_{m} J \\
\displaystyle\frac{\partial {J}}{\partial t}=\left(i \omega_a-\gamma_{a} / 2\right) J+i \sum_{m=1}^2 g_{m}F_m D, \\
\displaystyle\frac{\partial {D}}{\partial t}=-\gamma_{I}\left(D-D_{0}\right)+2 i \sum_{m=1}^2g_{m}\left(F^{*}_m J-J^{\ast} F_m\right)
\end{array}
\end{aligned}
\end{equation}
where $F$, $J$ and $D$ are proportional to the electric field $E_m$, the collective atomic polarization and the population difference respectively; $\gamma_I = \gamma_P + \gamma_D$; $\gamma_a = \gamma_P + \gamma_D + \gamma_E$; $D_0 = \frac{\gamma_P-\gamma_D}{\gamma_P+\gamma_D}$ is the pump parameter and $\widetilde{c}$ is the effective light velocity, and $\gamma_m$ is the photonic loss of the $m$th mode. 

Derivation of the Lugiato–Lefever equation from the Maxwell–Bloch equations (\ref{MBE}) for a case of a single photonic mode is given in \cite{Castelli2017} and discussed in other sources, e.g \cite{Lugiato2018}. In this work, we show the relationship between the maser soliton system and the LLE as based on derivation in 
 \cite{Castelli2017}. For simplicity we consider only one photonic mode, so that the Maxwell-Bloch equations in the frame associated with angular frequency $\omega_0$ can, by substituting $J\rightarrow iJ$, be written as:
 \begin{equation}
\begin{aligned}
	\label{MBE2}
	 \begin{array}{c}
\displaystyle \frac{\partial F}{\partial t} + \widetilde{c}\frac{\partial F}{\partial z}=-\frac{\gamma}{2}\Big[\left(1 + i \frac{2\Omega}{\gamma}\right) F - \frac{2g}{\gamma} J\Big] \\
\displaystyle\frac{\partial {J}}{\partial t}=-\frac{\gamma_a}{2}\Big[\left(1+i \frac{2\Omega_a}{\gamma_a}\right) J- \frac{2g}{\gamma_a}F D\Big], \\
\displaystyle\frac{\partial {D}}{\partial t}=-\gamma_{I}\Big[D-D_{0} +\frac{2 g}{\gamma_I}\left(F^{*} J+J^{\ast} F\right)\Big]
\end{array}
\end{aligned}
\end{equation}
where $\Omega = \omega_0- \omega$ and $\Omega_a = \omega_0- \omega_a$ are mode and atom detunings. By substituting $D\rightarrow D\chi$, $J\rightarrow J\xi$, $F\rightarrow F \zeta $: 
 \begin{equation}
\begin{aligned}
	\label{MBE3}
	 \begin{array}{c}
\displaystyle \frac{\partial F}{\partial t} + \widetilde{c}\frac{\partial F}{\partial z}=-\frac{\gamma}{2}\Big[\left(1 + i \frac{2\Omega}{\gamma}\right) F - \frac{2g\xi}{\gamma\zeta} J\Big] \\
\displaystyle\frac{\partial {J}}{\partial t}=-\frac{\gamma_a}{2}\Big[\left(1+i \frac{2\Omega_a}{\gamma_a}\right) J- \frac{2g\chi\zeta}{\gamma_a\xi}F D\Big], \\
\displaystyle\frac{\partial {D}}{\partial t}=-\gamma_{I}\Big[D-D_0/\chi + \frac{2g\xi\zeta}{\gamma_I\chi}\left(F^{*} J+J^{\ast} F\right)\Big]
\end{array}
\end{aligned}
\end{equation}
So, to match the notation in \cite{Castelli2017}, we have to choose $\frac{2g\xi}{\gamma\zeta} = -2C$, $\frac{2g\chi\zeta}{\gamma_a\xi}=1$ and $\frac{2g\xi\zeta}{\gamma_I\chi}=\frac{1}{2}$ giving us $\zeta=-\frac{1}{2g}\sqrt{\frac{\gamma_I\gamma_a}{2}}$, $\xi = \frac{C\gamma}{2g^2}\sqrt{\frac{\gamma_I\gamma_a}{2}}$ and $\chi = -\frac{C\gamma\gamma_a}{2g^2}$, giving the following system:
 \begin{equation}
\begin{aligned}
	\label{MBE4}
	 \begin{array}{c}
\displaystyle \frac{\partial F}{\partial t} + \widetilde{c}\frac{\partial F}{\partial z}=-\frac{\gamma}{2}\Big[\left(1 + i \theta\right) F +2CJ\Big] \\
\displaystyle\frac{\partial {J}}{\partial t}=-\frac{\gamma_a}{2}\Big[\left(1+i \Delta\right) J- F D\Big], \\
\displaystyle\frac{\partial {D}}{\partial t}=-\gamma_{I}\Big[D-D_0/\chi + \frac{1}{2}\left(F^{*} J+J^{\ast} F\right)\Big]
\end{array}
\end{aligned}
\end{equation}
where $\theta = \frac{2\Omega}{\gamma}$ and $\Delta =\frac{2\Omega_a}{\gamma_a} $ which is exactly the starting system for derivation of the LLE in \cite{Castelli2017}. The only additional parameter is the scaled population inversion parameter $D_0/\chi$ which is controlled in our experiment via the high frequency pump.
For the case $D_0 = \chi$ and a circular cavity of radius $R$, the corresponding LLE is written as:
\begin{equation}
\begin{split}
\frac{\partial}{\partial \bar{t}} \tilde{F}(\bar{t}, \varphi)=-\tilde{F}(\bar{t},& \varphi)-i \theta_{0} \tilde{F}(\bar{t}, \varphi) 
+i \eta \frac{\beta}{2} \frac{\partial^{2}}{\partial \varphi^{2}} \tilde{F}(\bar{t}, \varphi)\\&+i \eta|\tilde{F}(\bar{t}, \varphi)|^{2} \tilde{F}(\bar{t}, \varphi)
\end{split}
\end{equation}
where $\varphi$ is an angular coordinate, $\eta= -|\Delta|/\Delta$, $\widetilde{F}=\sqrt{\frac{2C}{|\Delta|^3}}F$, $\overline{t}=\frac{\gamma}{2}t$, $\theta_0 = \theta - \frac{2C}{\Delta}$ and the dispersion coefficient is 
 \begin{equation}
\beta = \frac{2C\gamma_a\widetilde{c}^2}{|\Omega_a|^3R^2}
\end{equation}

Although we have demonstrated that the present system is related to the LLE through its connection to the Maxwell-Bloch equations, its dynamics are far more complicated. This additional dynamics makes the system deviate from expected behaviour of typical LLE systems. 

\subsection{Distinguishing Features of the Maser Soliton System}
While in optics, a system may be represented by a pure 1D circular cavity, our sapphire resonator is a large 3D cavity with a more complicated wave structure involving some dynamics of two additional degrees of freedom. Its large spatial footprint leads to significant spin ensemble broadening. A more complicated level structure of the $\textrm{Fe}^{3+}$ ensemble induces additional decoherence effects through interaction with higher energy levels. The same is true for the high frequency pump mode exhibiting additional dynamics potentially involving decoherence effects; in optical counterparts involving only the Kerr nonlinearity this additional decoherence is avoided all together. Furthermore, apart from the target $\textrm{Fe}^{3+}$ ensemble, the crystal hosts additional spin ensembles (e.g. chromium) involving spin-spin interaction as well as interactions with nuclei spin which is the main source of local decoherence in our target spin system. Clearly, the system is highly susceptibility to external magnetic fluctuations due to the fact that it is based on a spin ensemble. Finally, there are two cavity modes involved which are both coupled to the same spin ensemble and thus can interact. Overall, all these additional effects give observed deviations from the picture expected from the LLE equation.

\section{Conclusion}
These observations provide a window into the spontaneous self-organization behaviours within this nonlinear many-body spin system, worthy of further study. The system may be related to the Lugiato-Lefever equation (a damped, driven, non-linear wave equation) for which temporal solitons on top of a continuous wave pump are a solution and which likewise describes the dynamics of Kerr frequency combs \cite{Lugiato2018}. Due to various inhomogeneities of the spin ensemble, and imperfections in the crystal itself, inhomogeneity and non-periodicity can be observed in the soliton trains. These temporal results are more thoroughly presented in the appendices. The production of this maser-based comb structure, and associated soliton pulse train, in a simple sapphire system has great potential for further investigation, both theoretically, into the behaviour of the dissipative masing system, and into the potential applications of this comb, for example in precision spectroscopy~\cite{Diddams2007}, tunable narrowband microwave signal generation~\cite{Newbury2011} or reconfigurable RF filtering~\cite{Savchenkov2008}. Detailed characterization of the soliton evolution and comb structure will be undertaken in the future.

\section*{Acknowledgements}
This research was supported by the ARC Centre of Excellence for Engineered Quantum Systems (EQUS, CE170100009) and the ARC Centre of Excellence for Dark Matter Particle Physics (CDM, CE200100008).\hspace{10pt}

\appendix

\section{Results at 4~K} 
\begin{figure*}[h]
\begin{minipage}{.45\textwidth}
\centering
\subfloat[]{\label{}\includegraphics[scale=0.3]{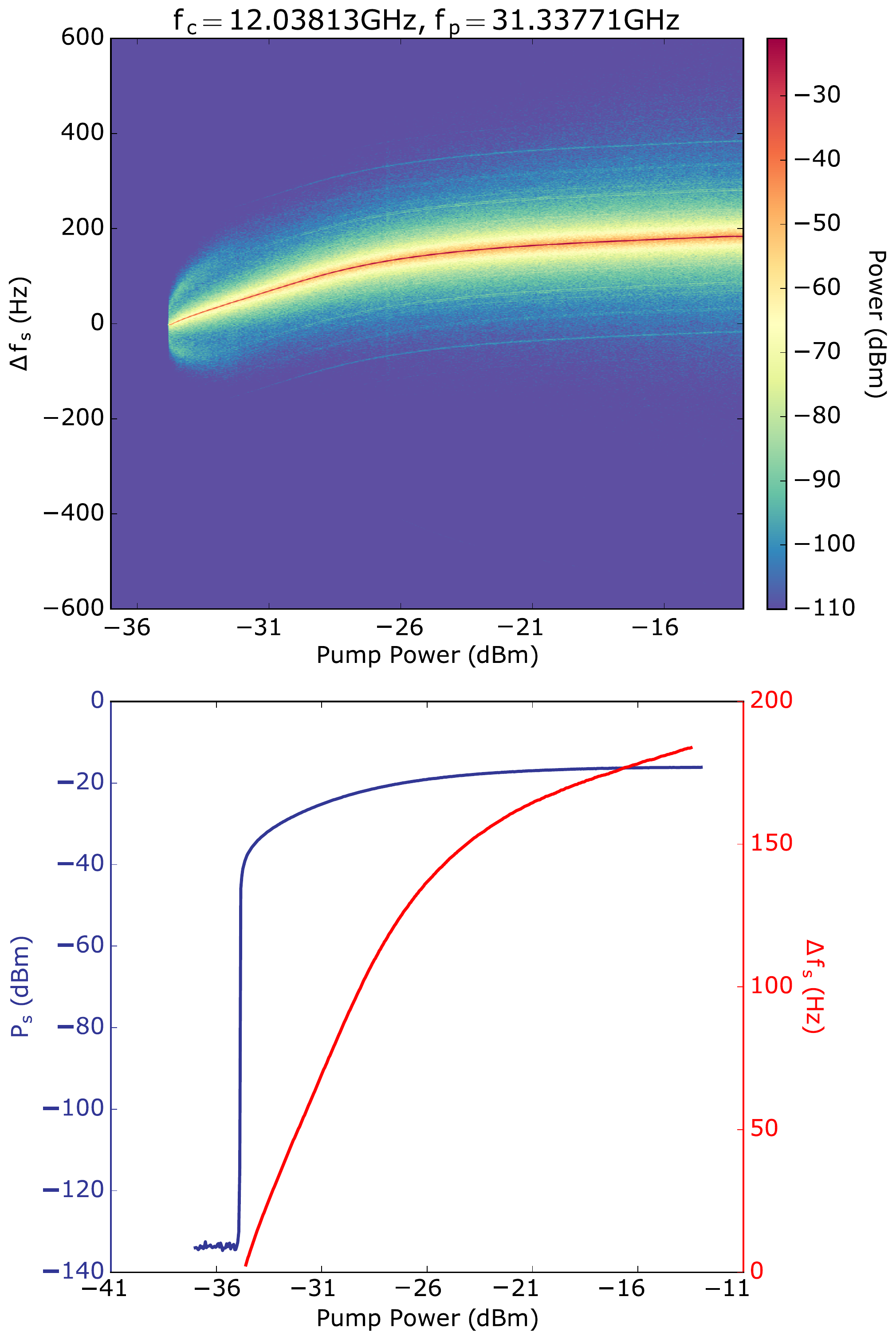}}
\end{minipage}%
\begin{minipage}{.45\textwidth}
\centering
\subfloat[]{\label{}\includegraphics[scale=0.3]{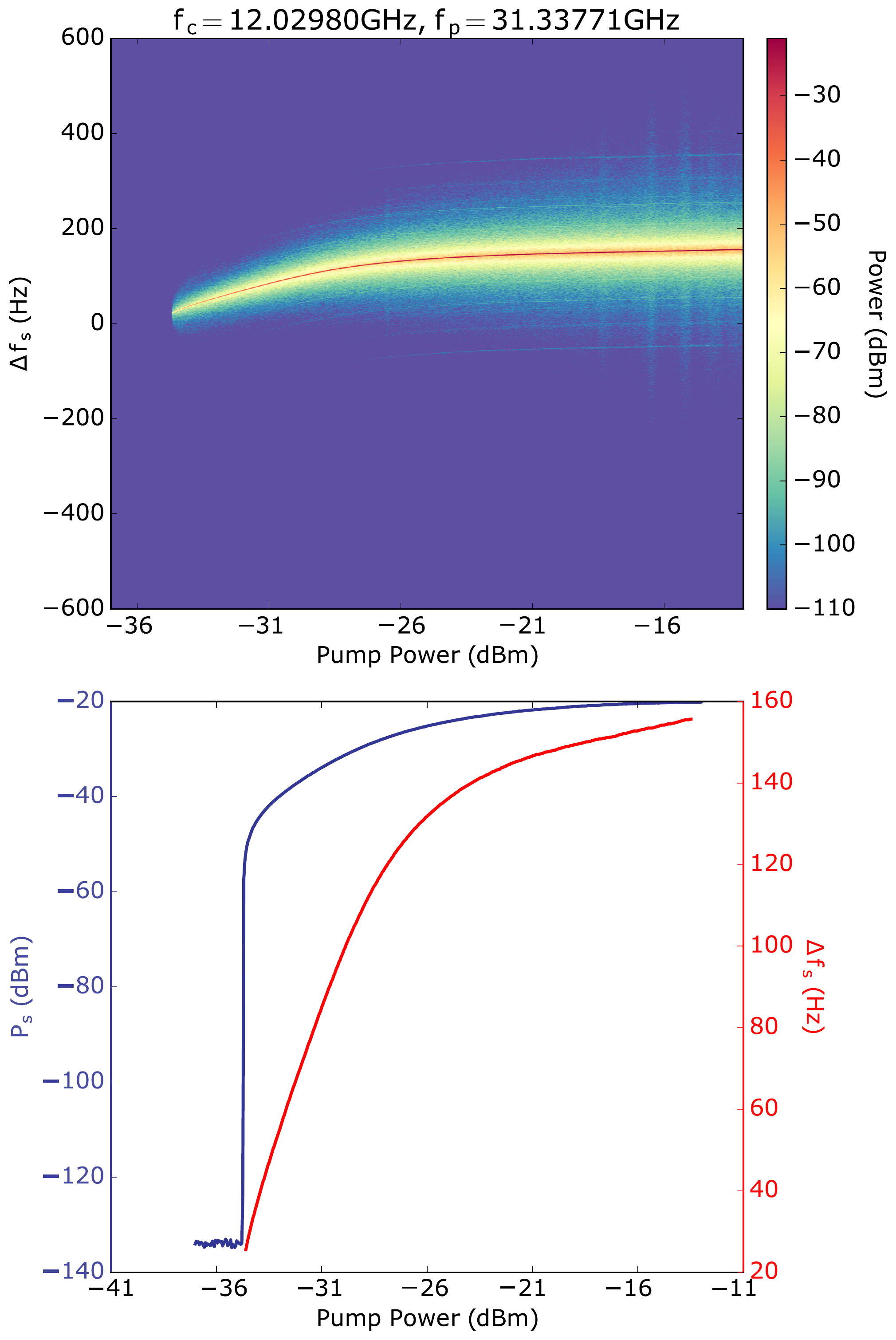}}
\end{minipage}%
\caption{Pumping on the first pump mode at an array of pump powers for modes A (a) and B (b). Measurements taken at 4~K. Power at signal output is gauged by the color bar and the masing peak can be tracked relative to the centre of the measurement spectrum ($\Delta f_{s}$). The bottom plots isolate the frequency shift  ($\Delta f_{s}$, red) and maximum power ($P_s$, purple) of the maser peaks, depending on incident pump power.}
\label{4KP}
\end{figure*}

\begin{figure*}[t] 
\begin{minipage}{.33\linewidth}
\centering
\subfloat[]{\label{main:b}\includegraphics[scale=.28]{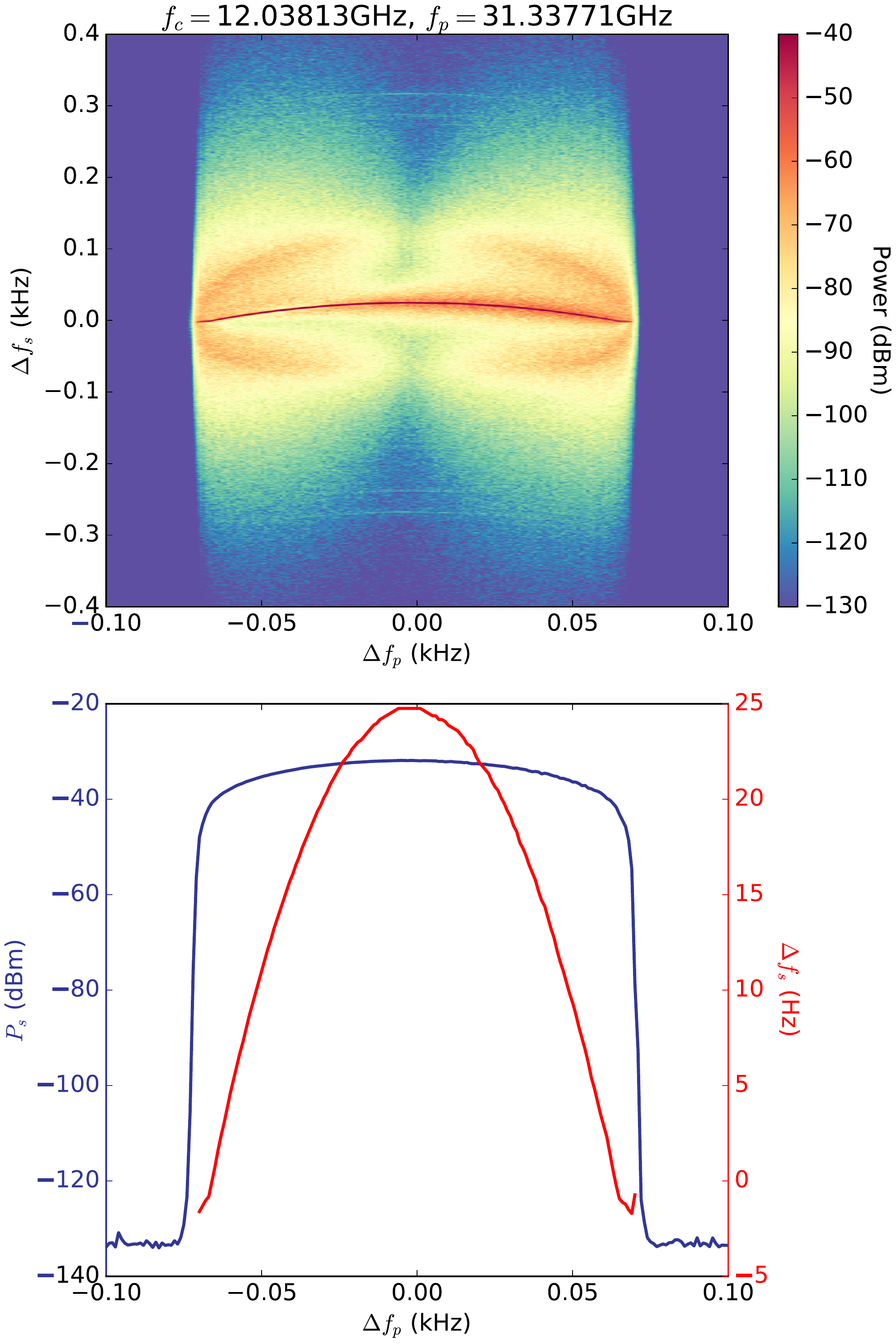}}
\end{minipage}
\begin{minipage}{.33\linewidth}
\centering
\subfloat[]{\label{}\includegraphics[scale=.28]{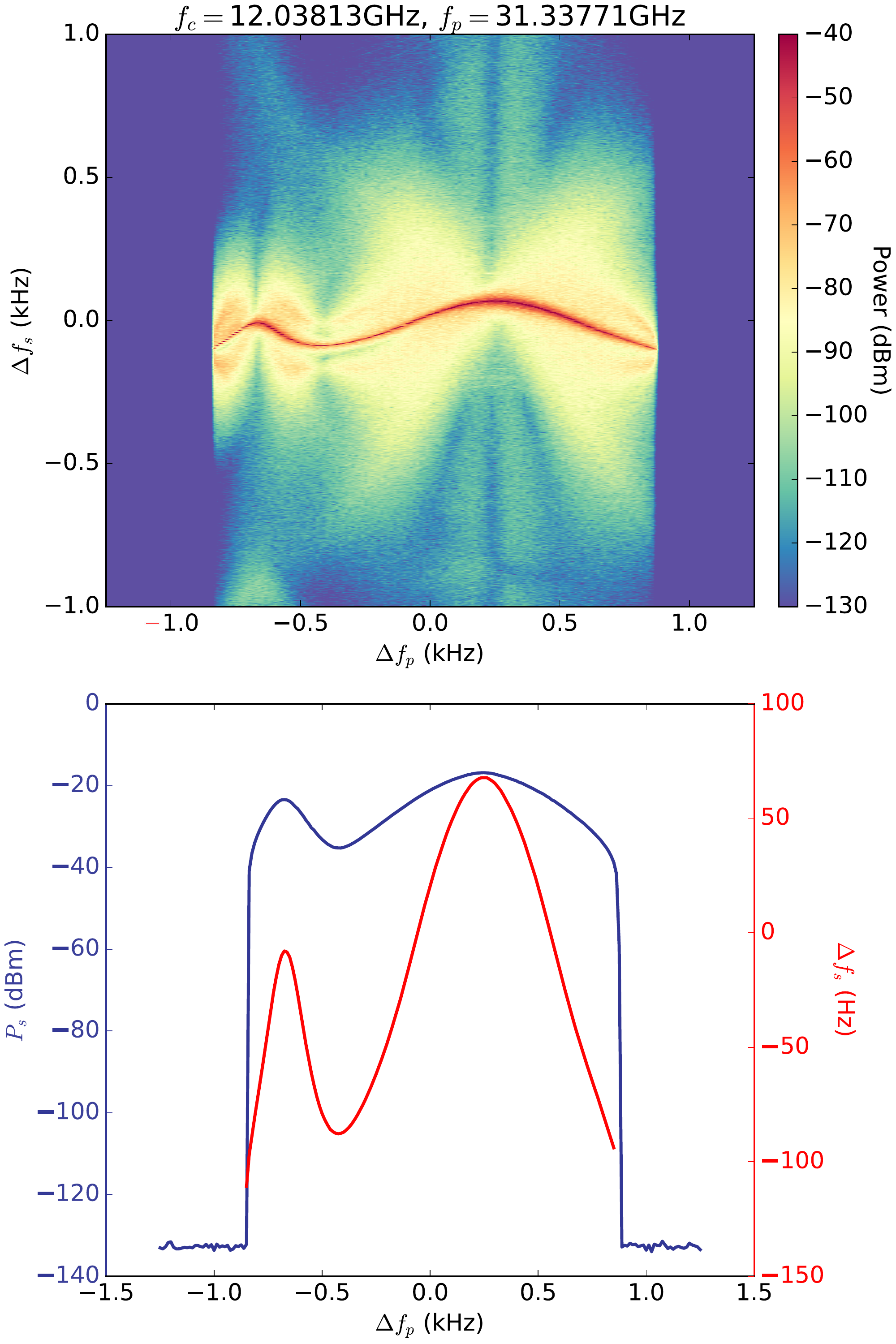}}
\end{minipage}%
\begin{minipage}{.33\linewidth}
\centering
\subfloat[]{\label{}\includegraphics[scale=.28]{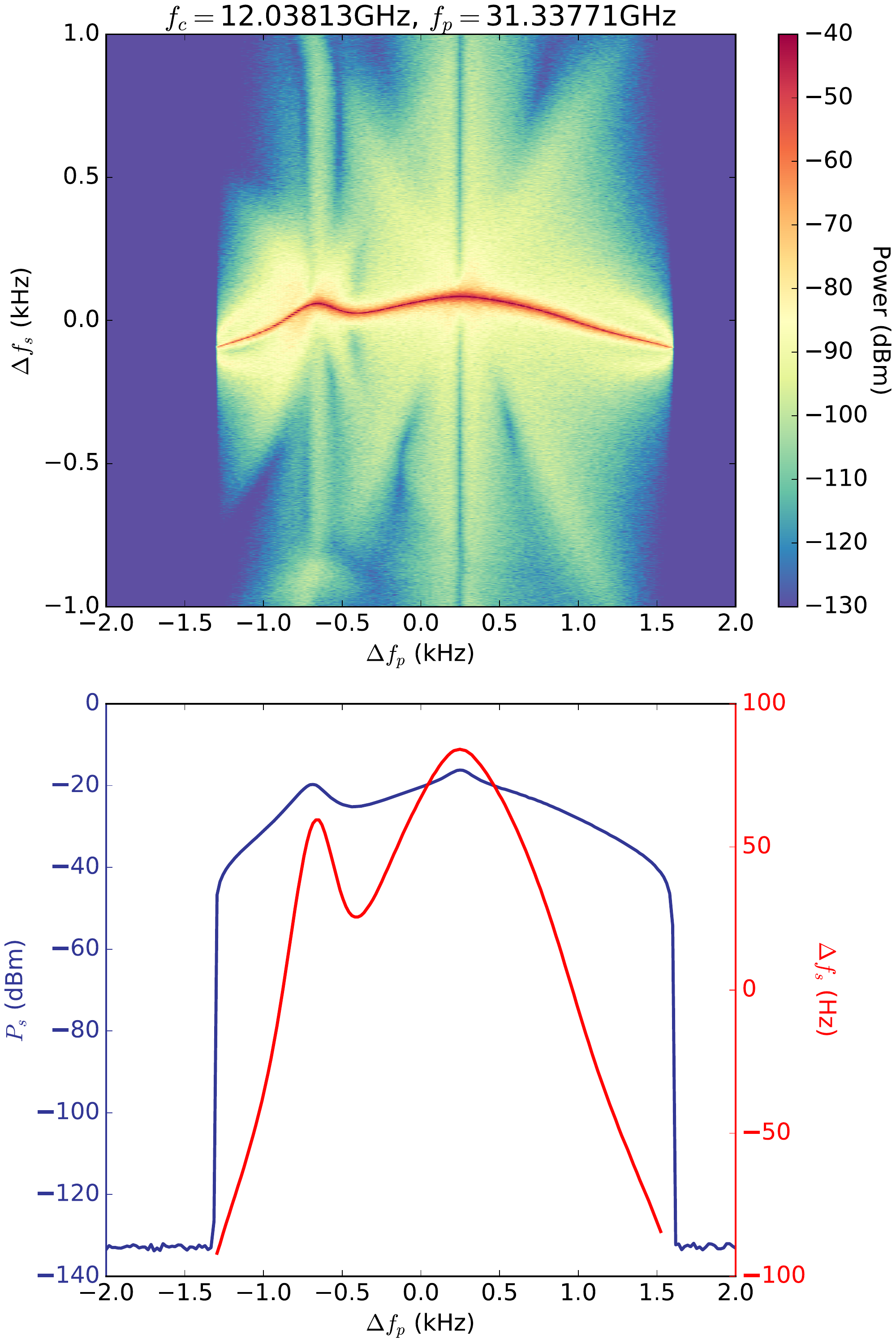}}
\end{minipage}%

\begin{minipage}{.33\linewidth}
\vspace{4 mm}
\centering
\subfloat[]{\label{main:b}\includegraphics[scale=.28]{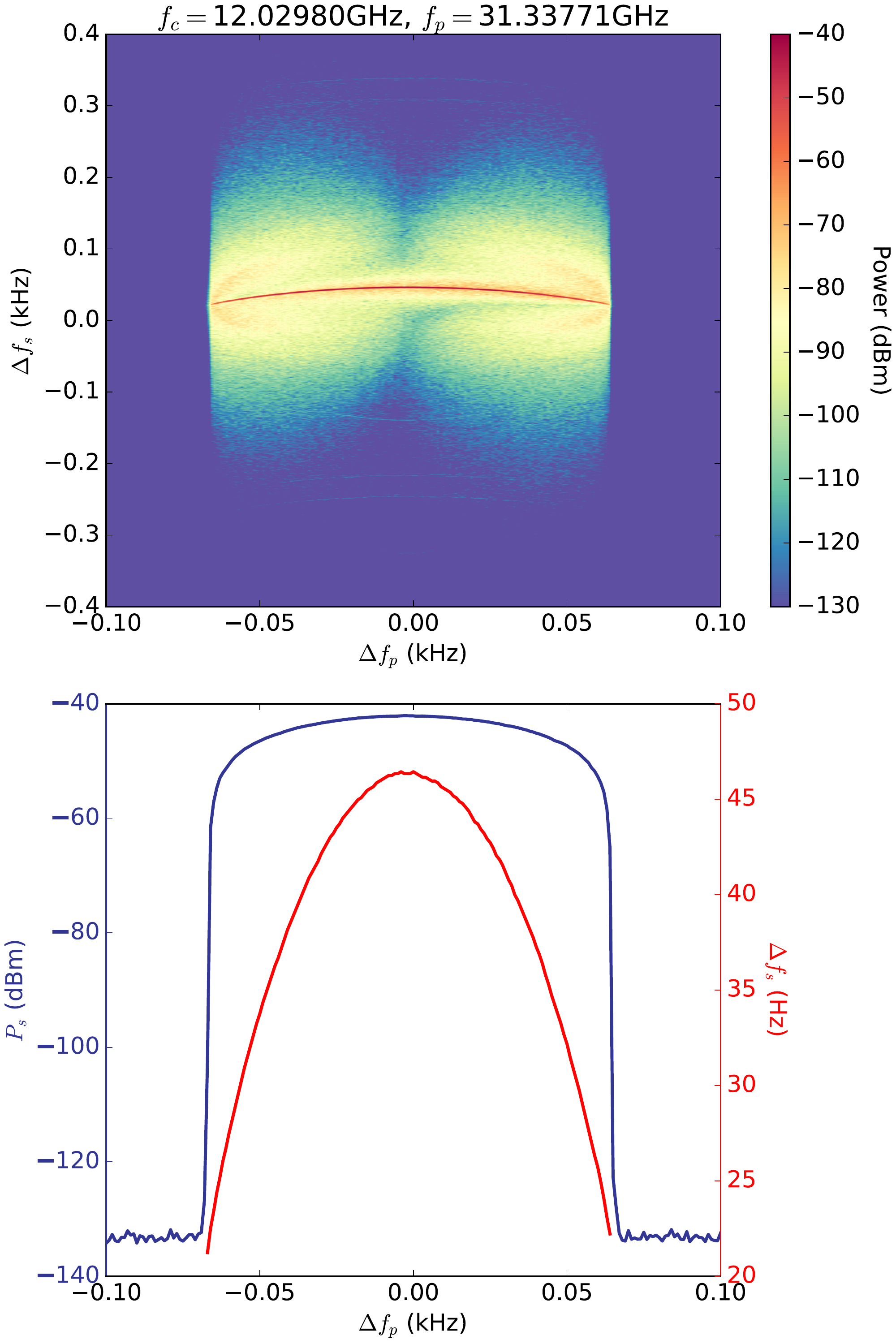}}
\end{minipage}
\begin{minipage}{.33\linewidth}
\vspace{4 mm}
\centering
\subfloat[]{\label{}\includegraphics[scale=.28]{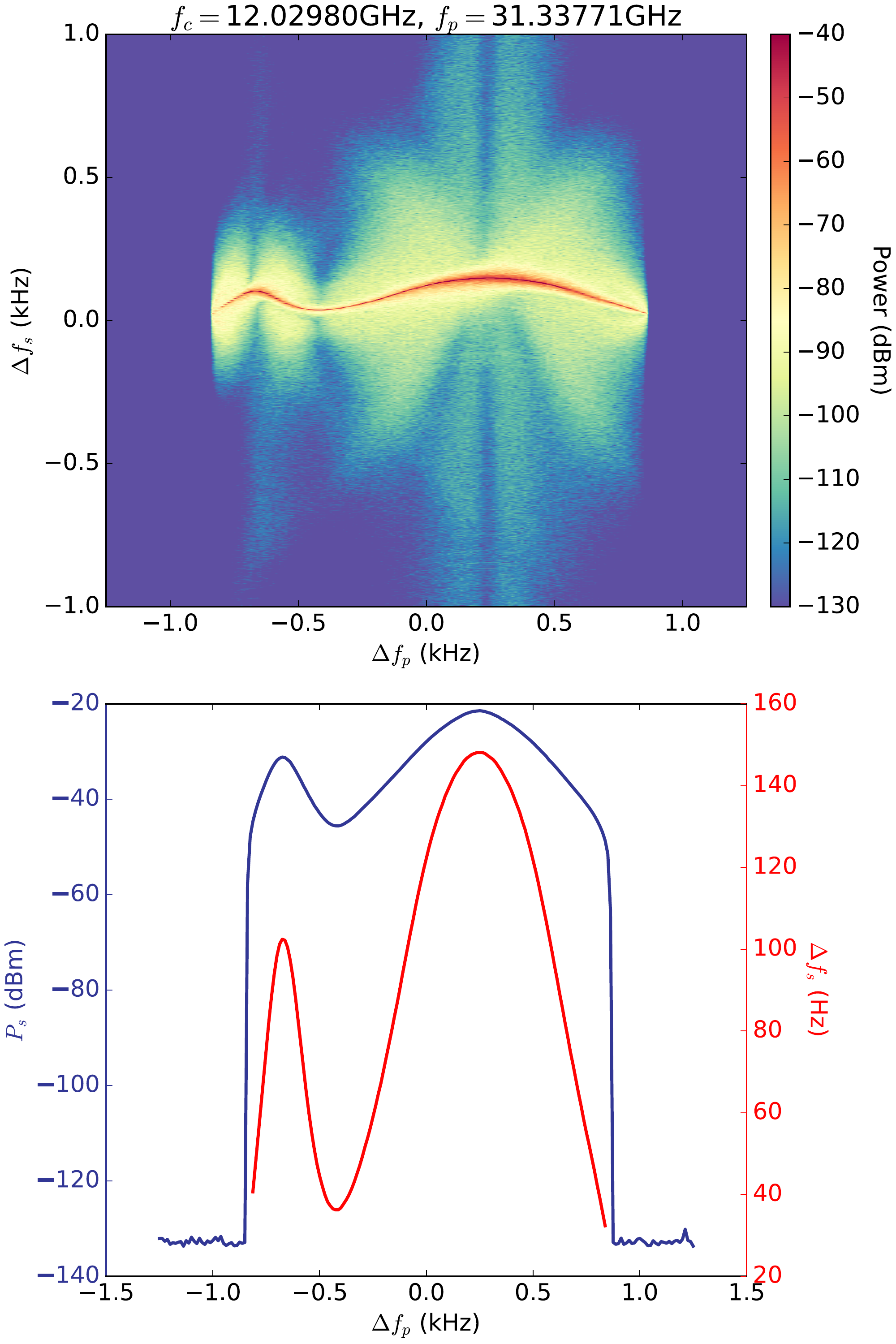}}
\end{minipage}%
\begin{minipage}{.33\linewidth}
\vspace{4 mm}
\centering
\subfloat[]{\label{}\includegraphics[scale=.28]{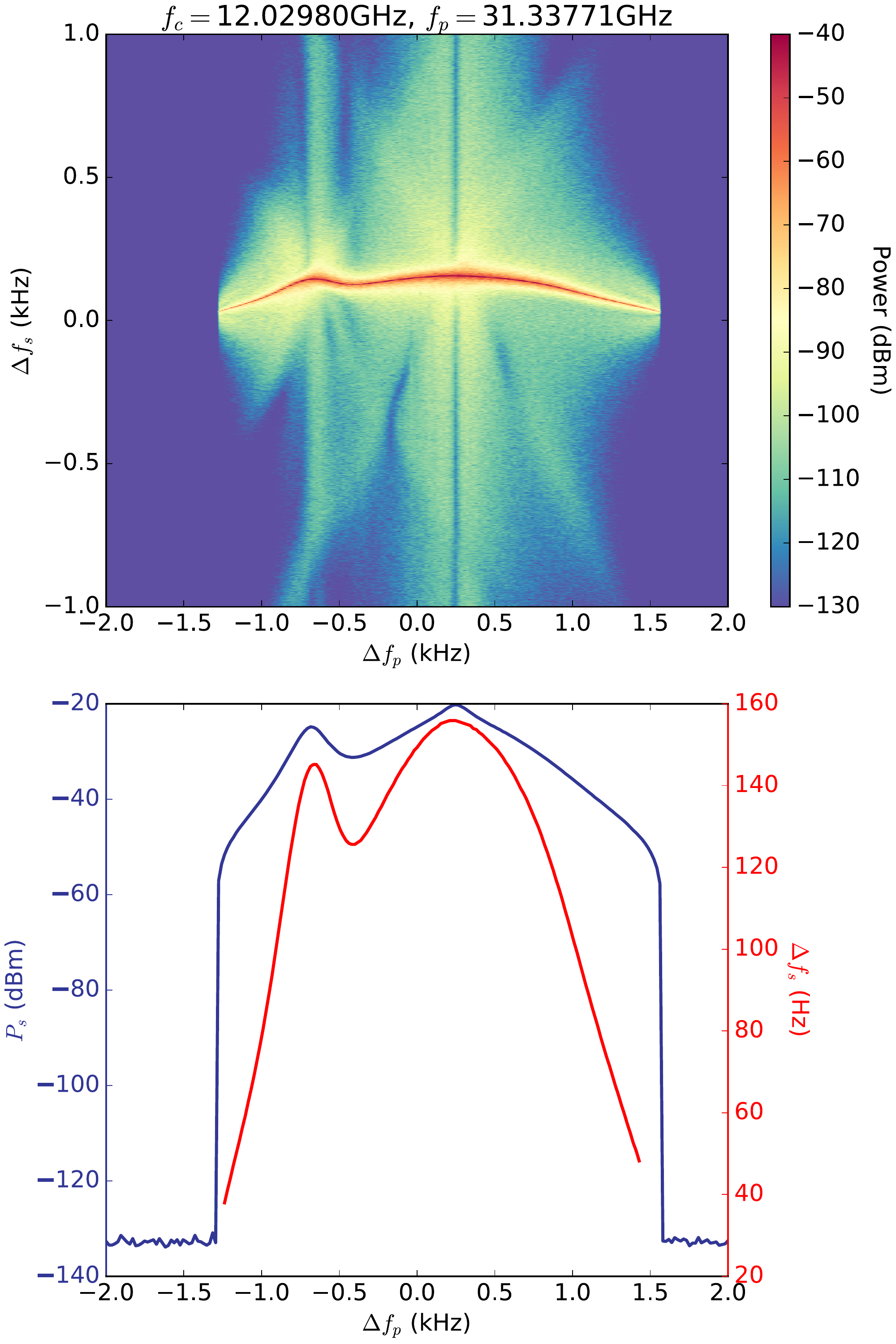}}
\end{minipage}%
\vspace{-2 mm}
\caption{Pumping on the first pump mode at an array of pump frequencies about $f_p$ for modes A (a,b,c) and B (d,e,f) with pump power set to -33~dBm, -20~dBm and -13~dBm respectively. Measurements taken at 4~K. Power at signal output is gauged by the color bar and the masing peak can be tracked relative to the centre of the measurement spectrum ($\Delta f_{s}$). The bottom plots isolate the frequency shift  ($\Delta f_{s}$, red) and maximum power ($P_s$, purple) of the maser peaks, depending on incident pump frequency.}
\label{4KAandB}
\end{figure*}

Fig.~\ref{4KP} presents the power-dependence of masing, for modes A and B. Unlike the masing results at mK, shown in the main paper, there is no two peak structure present, and although there are modulation sidebands, a fine frequency comb is not produced as in the mK case. Fig~\ref{4KAandB} presents the response of masing on modes A and B when the pump frequency is swept. As in the mK case, the doublet structure of the pump mode is evident.

\section{Further Results at mK}
\begin{figure*}[h]
\begin{minipage}{.33\linewidth}
\centering
\subfloat[]{\label{}\includegraphics[scale=.28]{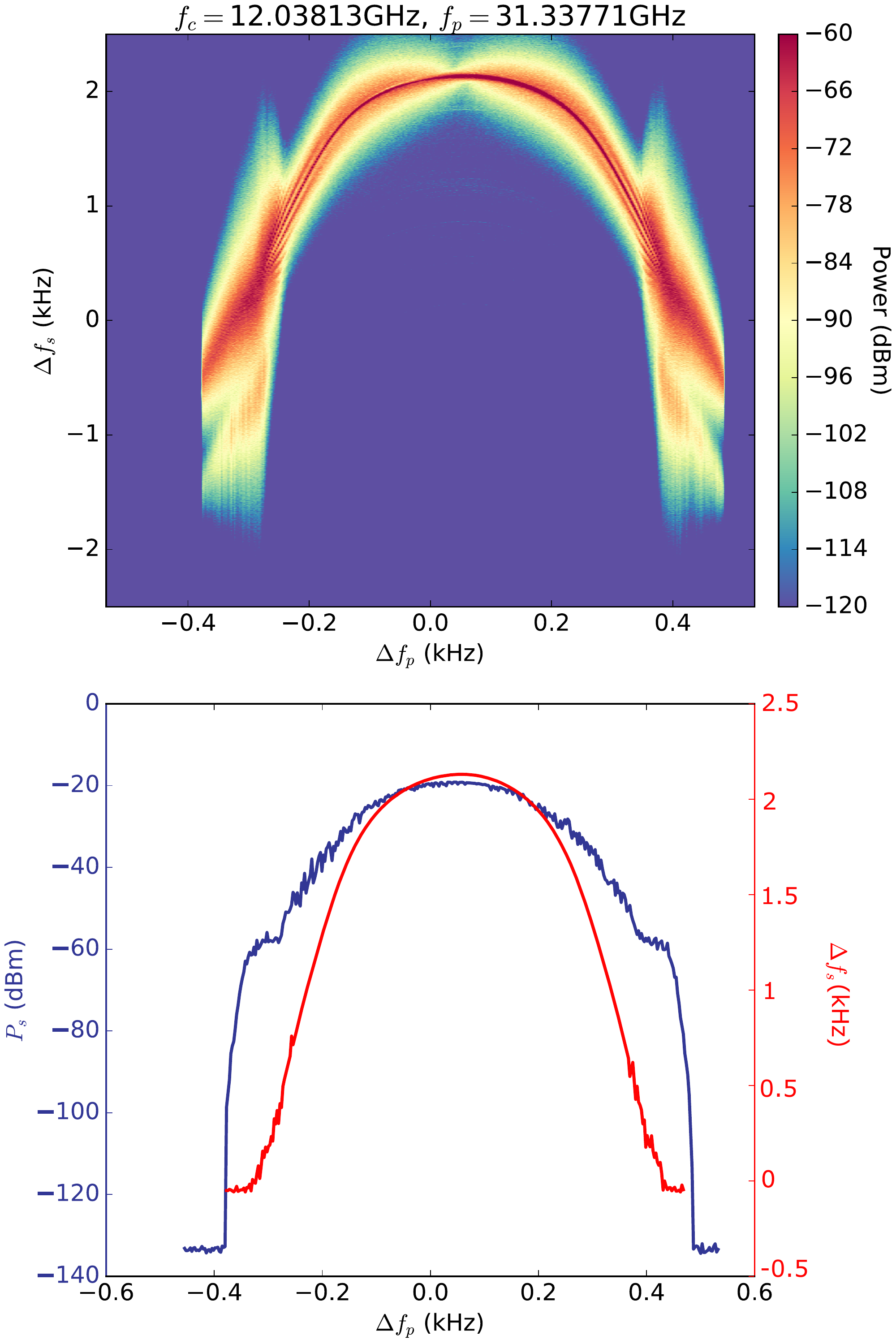}}
\end{minipage}%
\begin{minipage}{.33\linewidth}
\centering
\subfloat[]{\label{}\includegraphics[scale=.28]{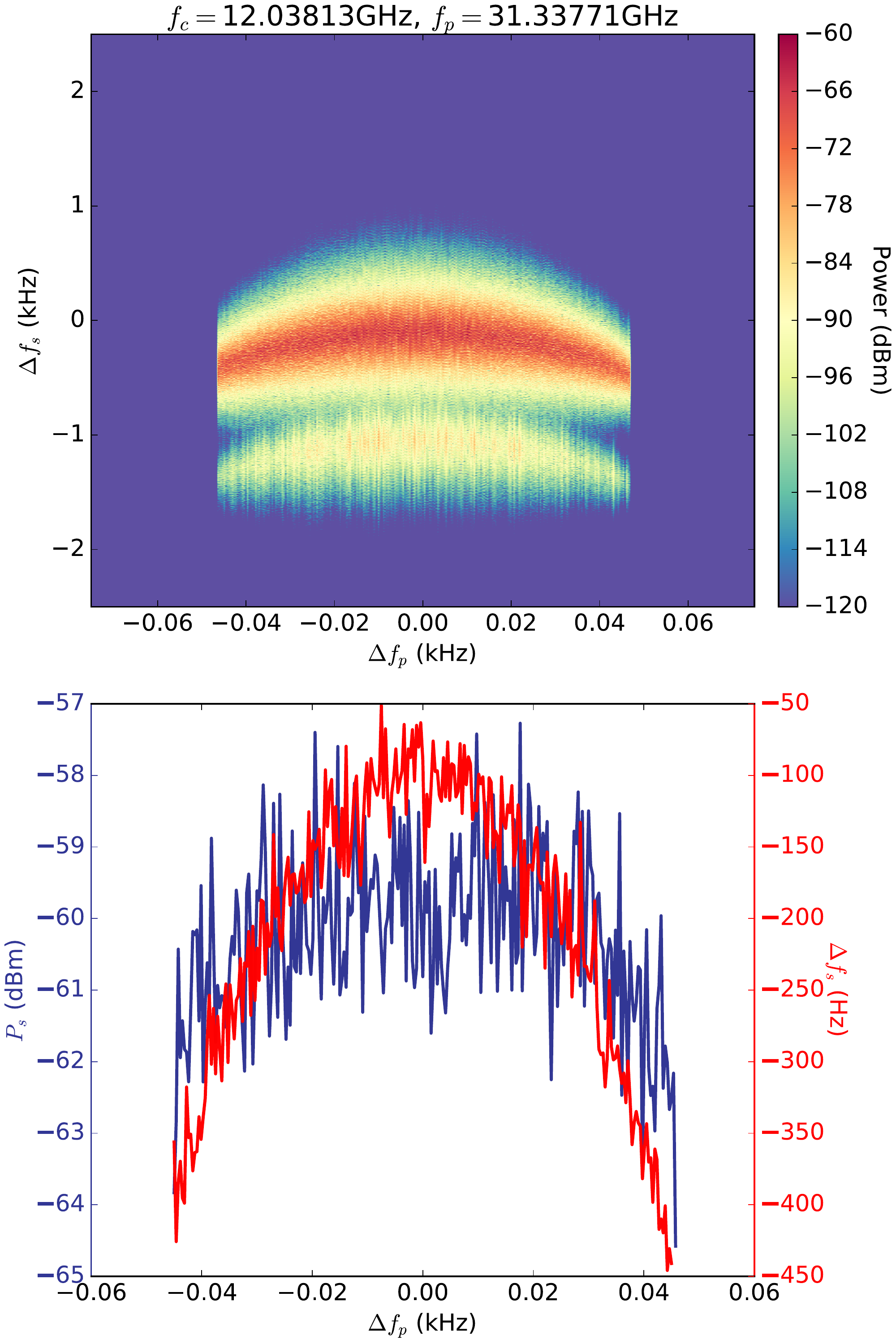}}
\end{minipage}%
\begin{minipage}{.33\linewidth}
\centering
\subfloat[]{\label{main:b}\includegraphics[scale=.28]{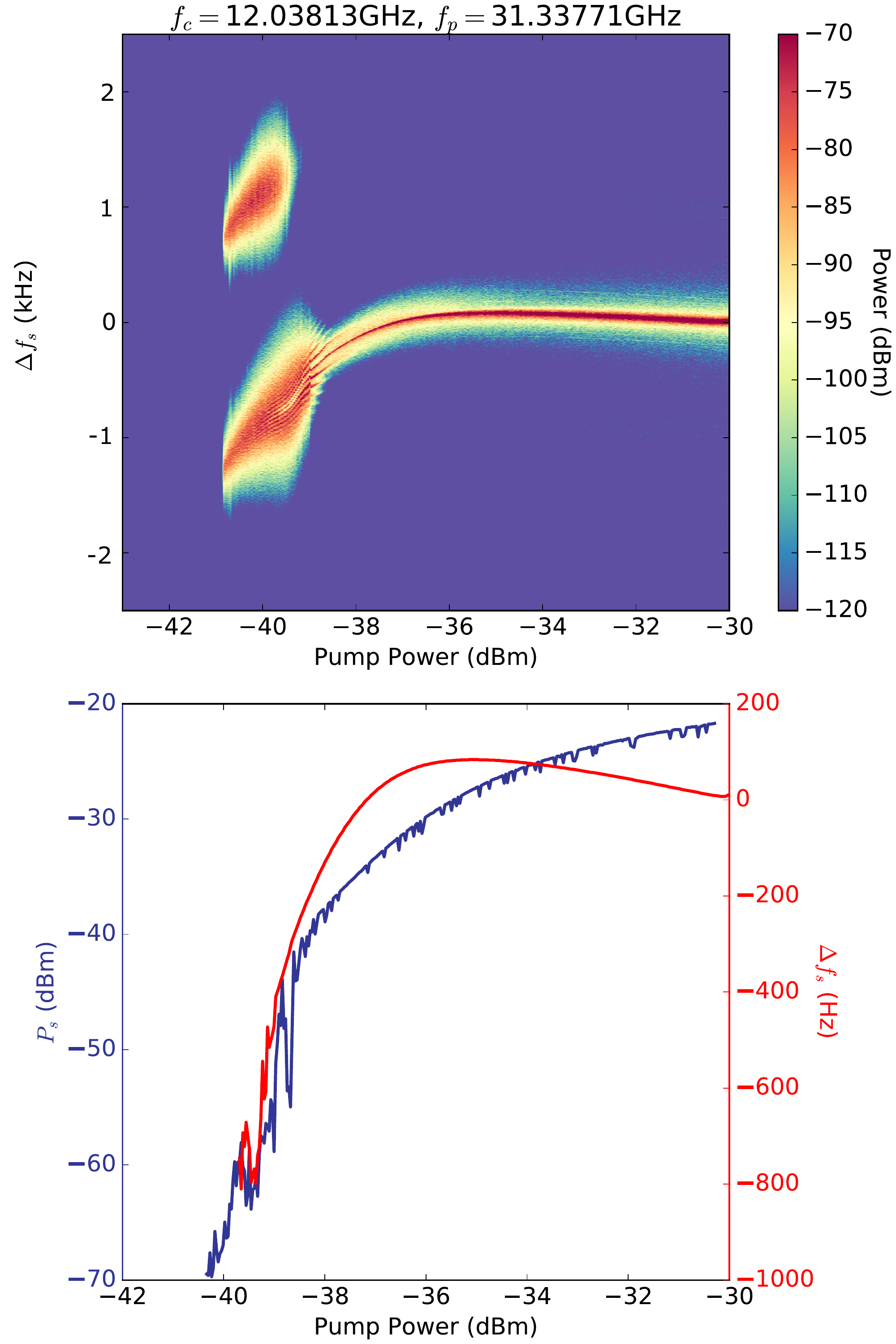}}
\end{minipage}

\begin{minipage}{.33\linewidth}
\vspace{4 mm}
\centering
\subfloat[]{\label{main:b}\includegraphics[scale=.28]{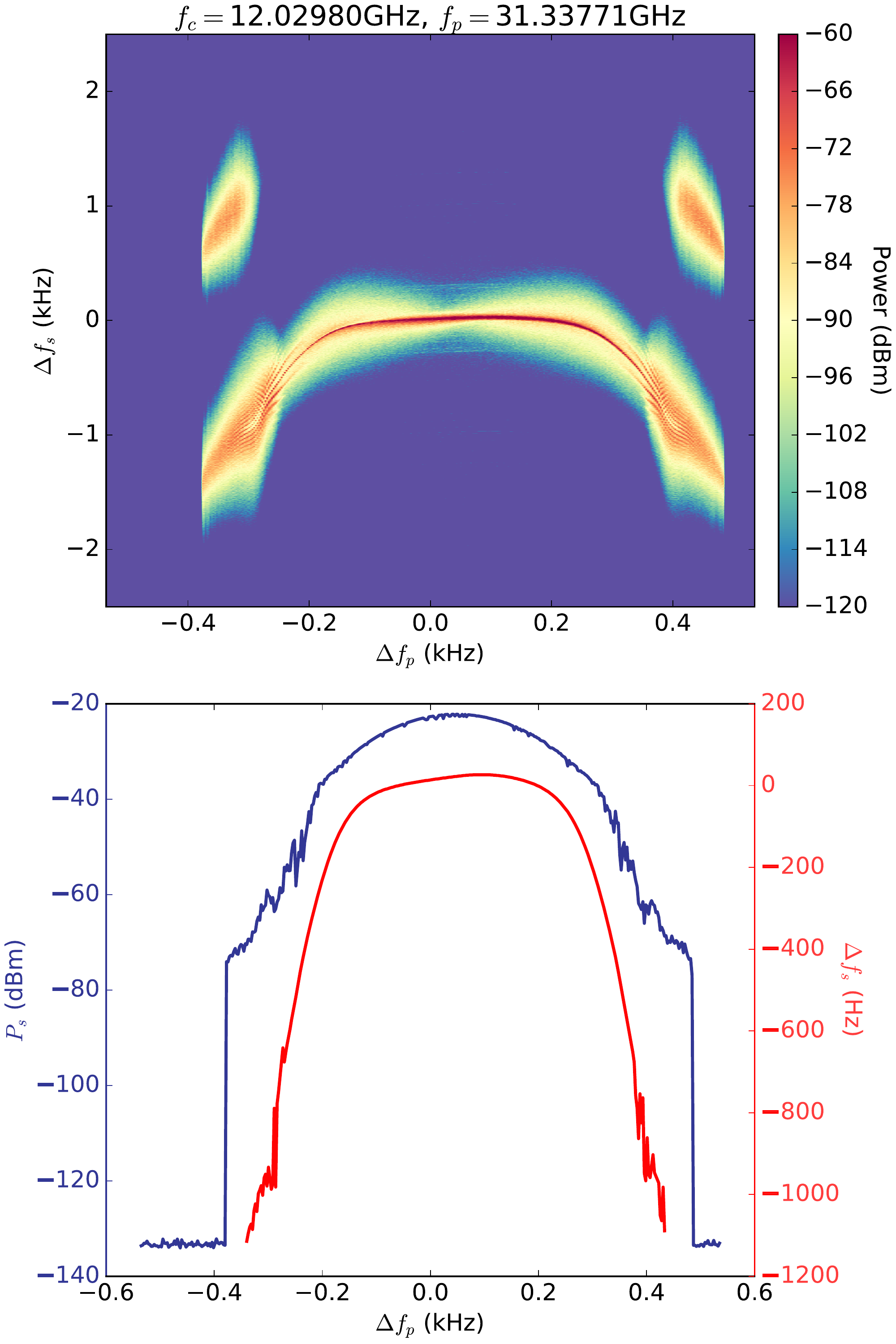}}
\end{minipage}
\begin{minipage}{.33\linewidth}
\vspace{4 mm}
\centering
\subfloat[]{\label{}\includegraphics[scale=.28]{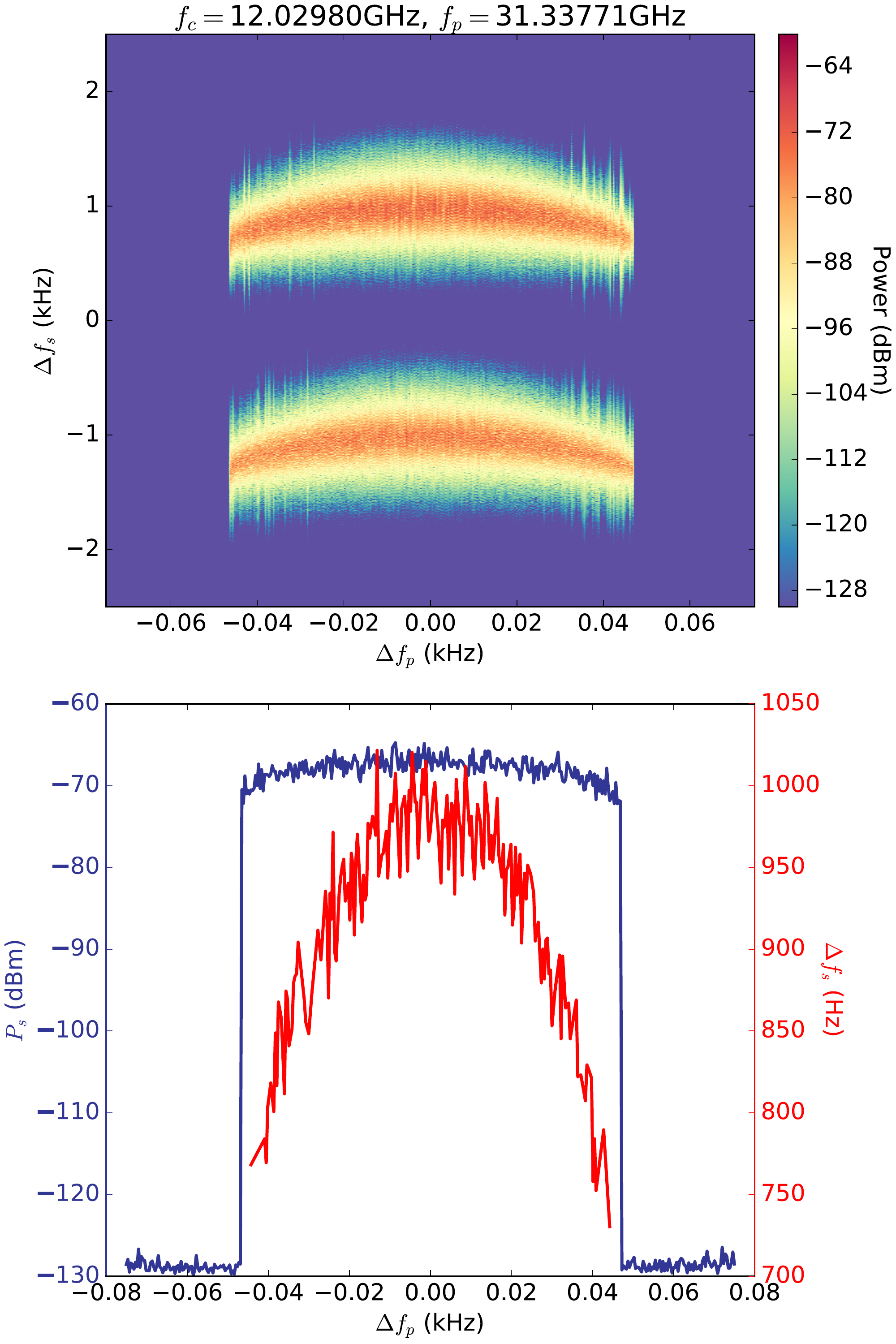}}
\end{minipage}%
\begin{minipage}{.33\linewidth}
\vspace{4 mm}
\centering
\subfloat[]{\label{}\includegraphics[scale=.28]{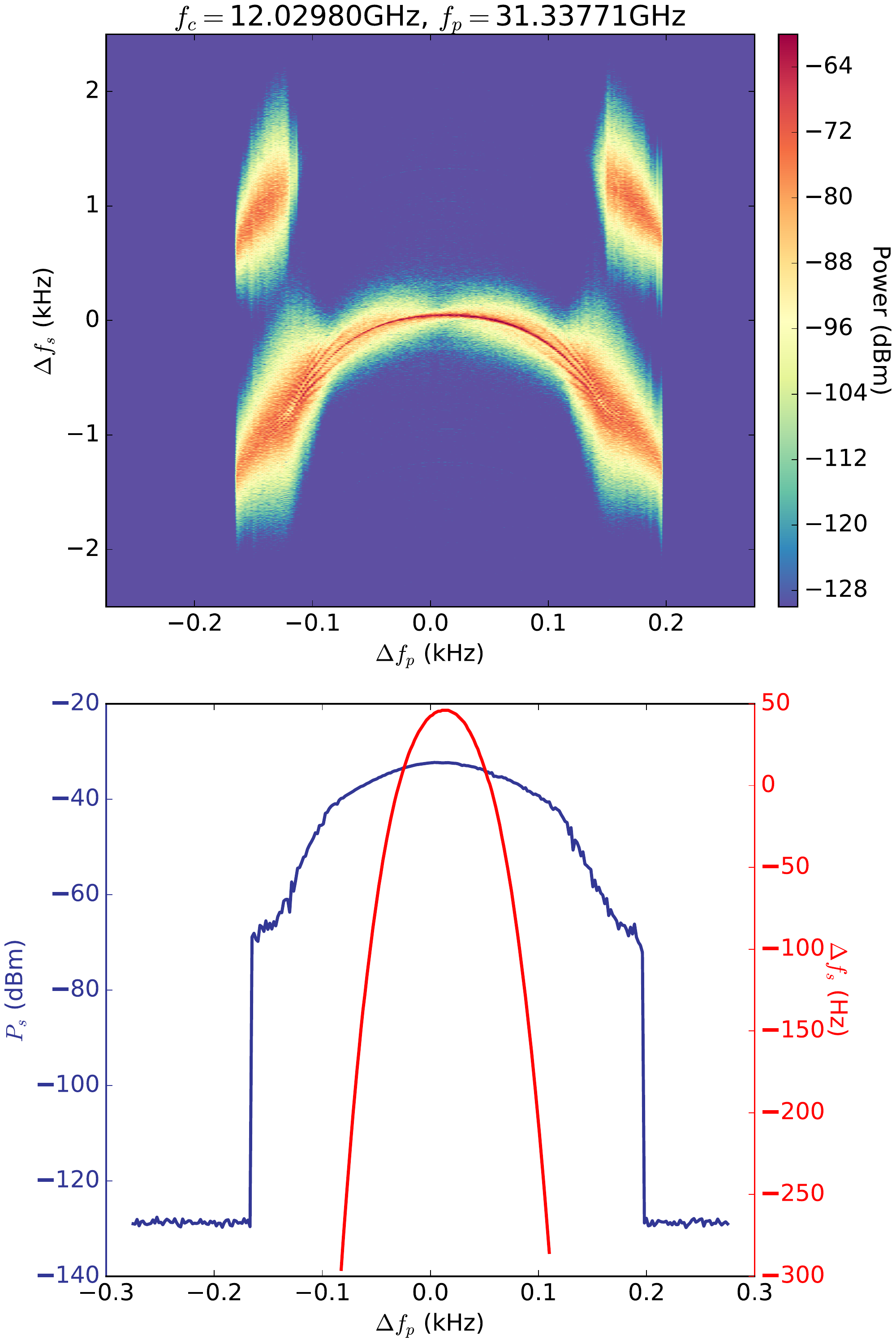}}
\end{minipage}%
\caption{Pumping on the first pump mode at an array of pump frequencies about $f_p$ for modes A (a,b) and B (d,e,f) with pump power set to -31~dBm, -41~dBm and -37~dBm (mode B only) respectively. Measurements taken at mK. Power at signal output is gauged by the color bar and the masing peak can be tracked relative to the centre of the measurement spectrum ($\Delta f_{s}$). The bottom plots isolate the frequency shift  ($\Delta f_{s}$, red) and maximum power ($P_s$, purple) of the maser peaks, depending on input pump frequency. The dependence of mode B upon pump power is represented in (c), but neglected for mode A, as this is shown in the main paper. }
\label{mKExtra1}
\end{figure*}

\begin{figure*}[h]
\begin{minipage}{.33\linewidth}
\centering
\subfloat[]{\label{}\includegraphics[scale=.28]{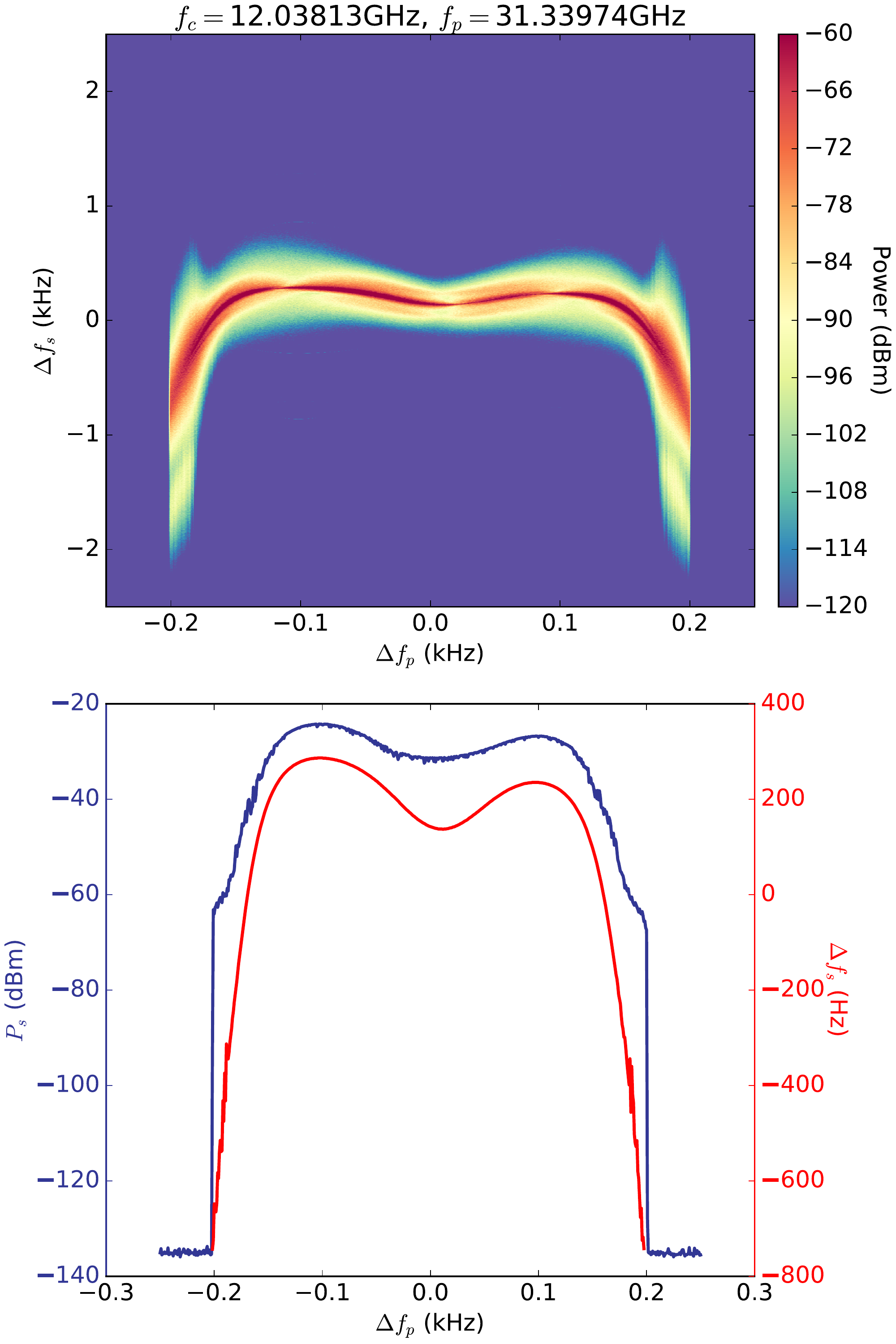}}
\end{minipage}%
\begin{minipage}{.33\linewidth}
\centering
\subfloat[]{\label{}\includegraphics[scale=.28]{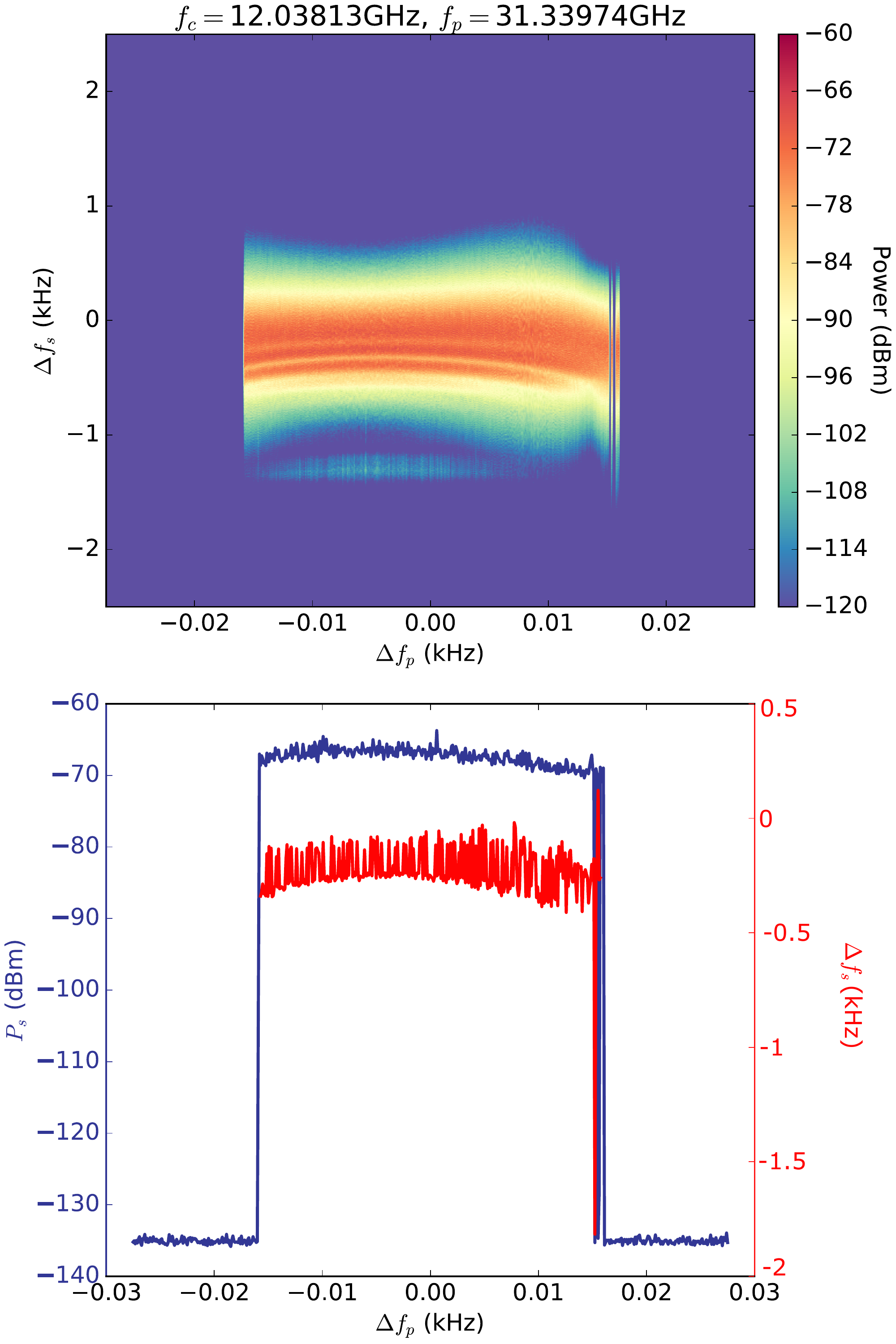}}
\end{minipage}%
\begin{minipage}{.33\linewidth}
\centering
\subfloat[]{\label{main:b}\includegraphics[scale=.28]{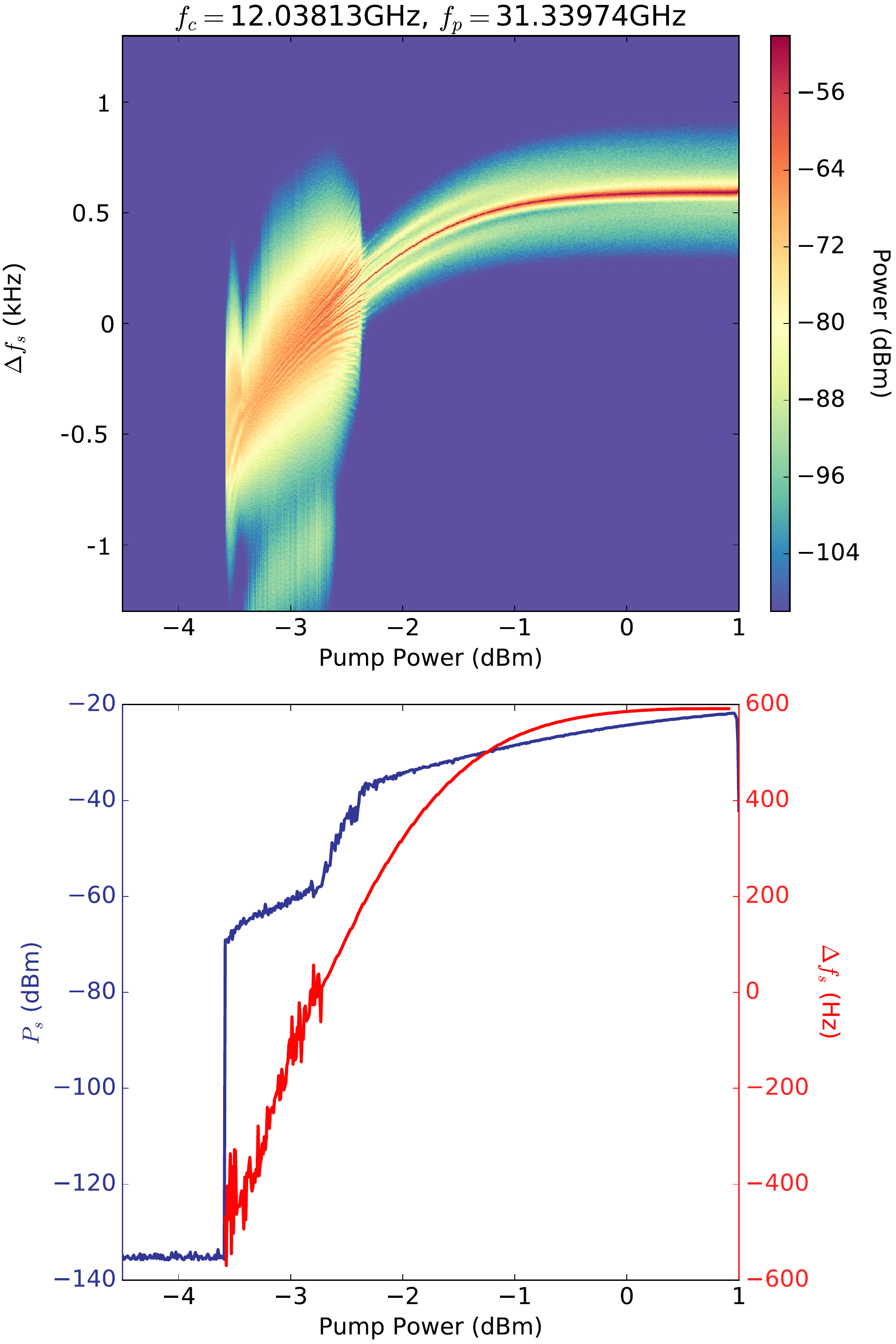}}
\end{minipage}

\begin{minipage}{.33\linewidth}
\vspace{4 mm}
\centering
\subfloat[]{\label{main:b}\includegraphics[scale=.28]{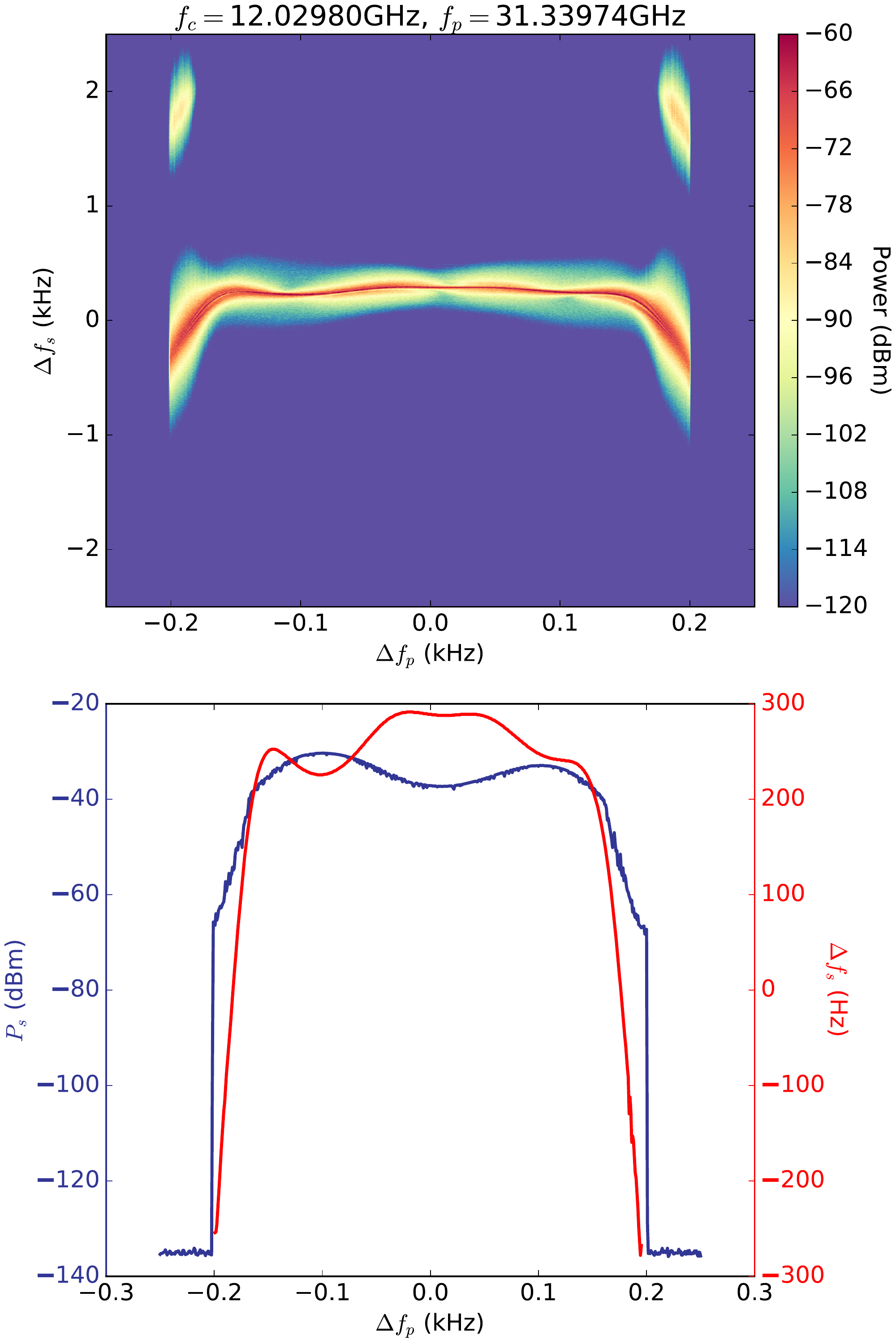}}
\end{minipage}
\begin{minipage}{.33\linewidth}
\vspace{4 mm}\centering
\subfloat[]{\label{}\includegraphics[scale=.28]{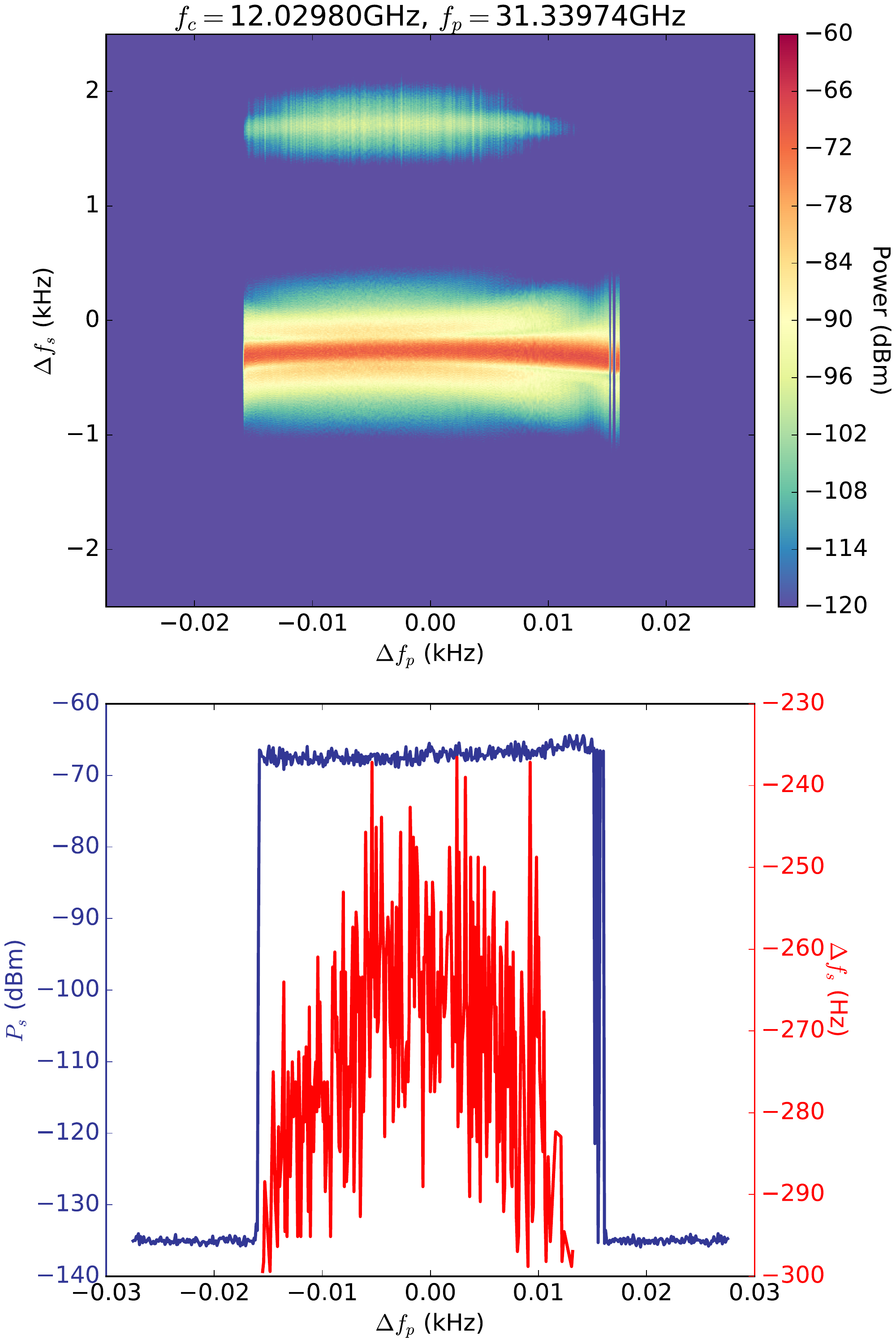}}
\end{minipage}%
\begin{minipage}{.33\linewidth}
\vspace{4 mm}
\centering
\subfloat[]{\label{}\includegraphics[scale=.28]{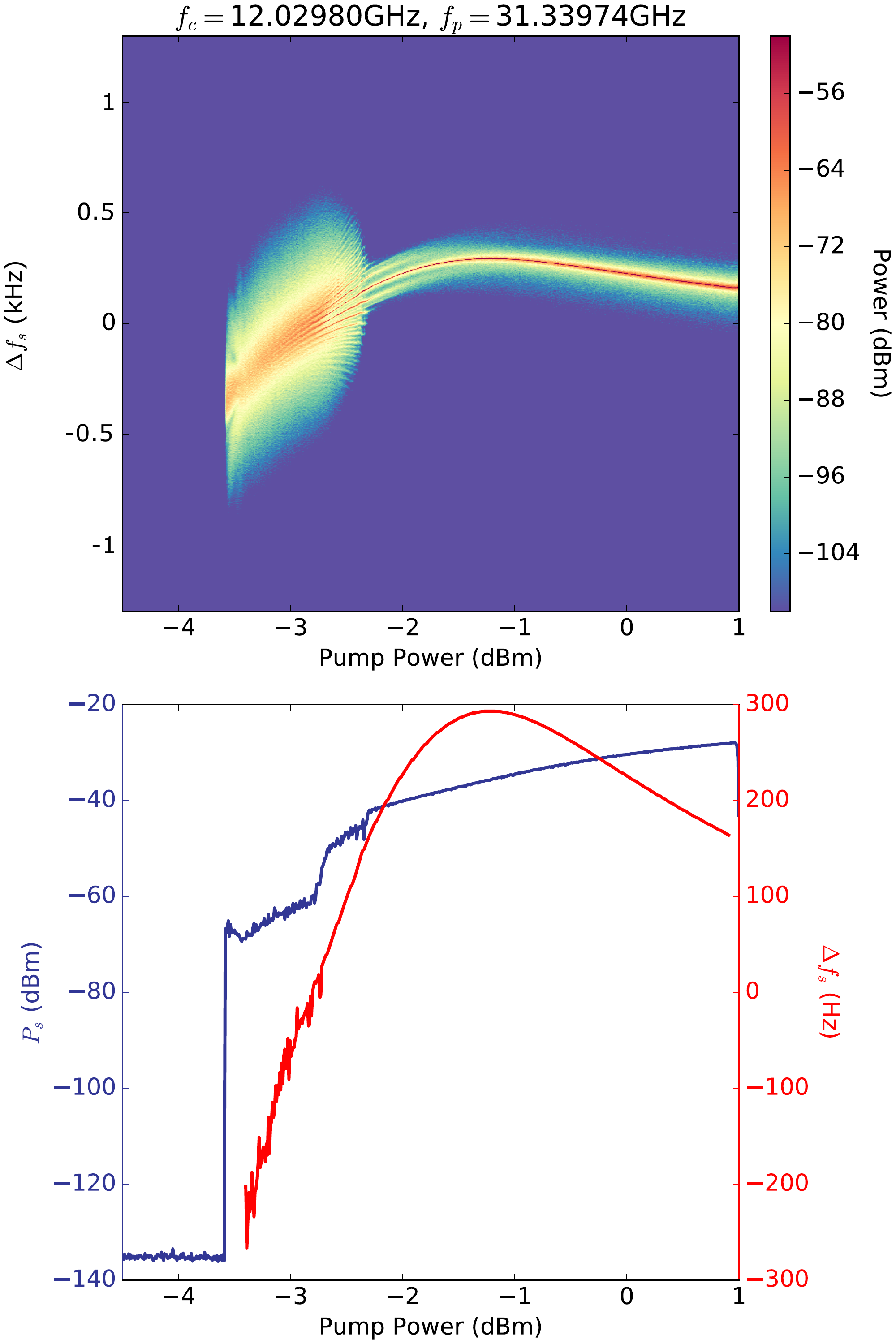}}
\end{minipage}%
\caption{Pumping on the second pump mode at an array of pump frequencies about $f_p$ for modes A (a,b) and B (d,e) with pump power set to -31~dBm and -34.5~dBm respectively. Measurements taken at mK. Power at signal output is gauged by the color bar and the masing peak can be tracked relative to the centre of the measurement spectrum ($\Delta f_{s}$). The bottom plots isolate the frequency shift  ($\Delta f_{s}$, red) and maximum power ($P_s$, purple) of the maser peaks, depending on input pump frequency. The dependence of modes A and B upon pump power is also represented ((c) and (f)). }
\label{mKExtra2}
\end{figure*}

Fig.~\ref{mKExtra1} and~\ref{mKExtra2} present masing observed at mK, that is, between 16 mK and 50 mK after transient heating, over a range of pump frequencies and at different incident powers. Fig.~\ref{mKExtra1} presents results when pump mode 1 was pumped and Fig.~\ref{mKExtra2} presents results when pump mode 2 was pumped. Fig.~\ref{FIG11} and Fig.~\ref{FIG12} display the time series soliton output.
\vspace{5 mm}

\vspace{-3 mm}
\begin{figure*}[hbt]
\begin{minipage}{.33\linewidth}
\centering
\subfloat[]{\label{main:b}\includegraphics[scale=.28]{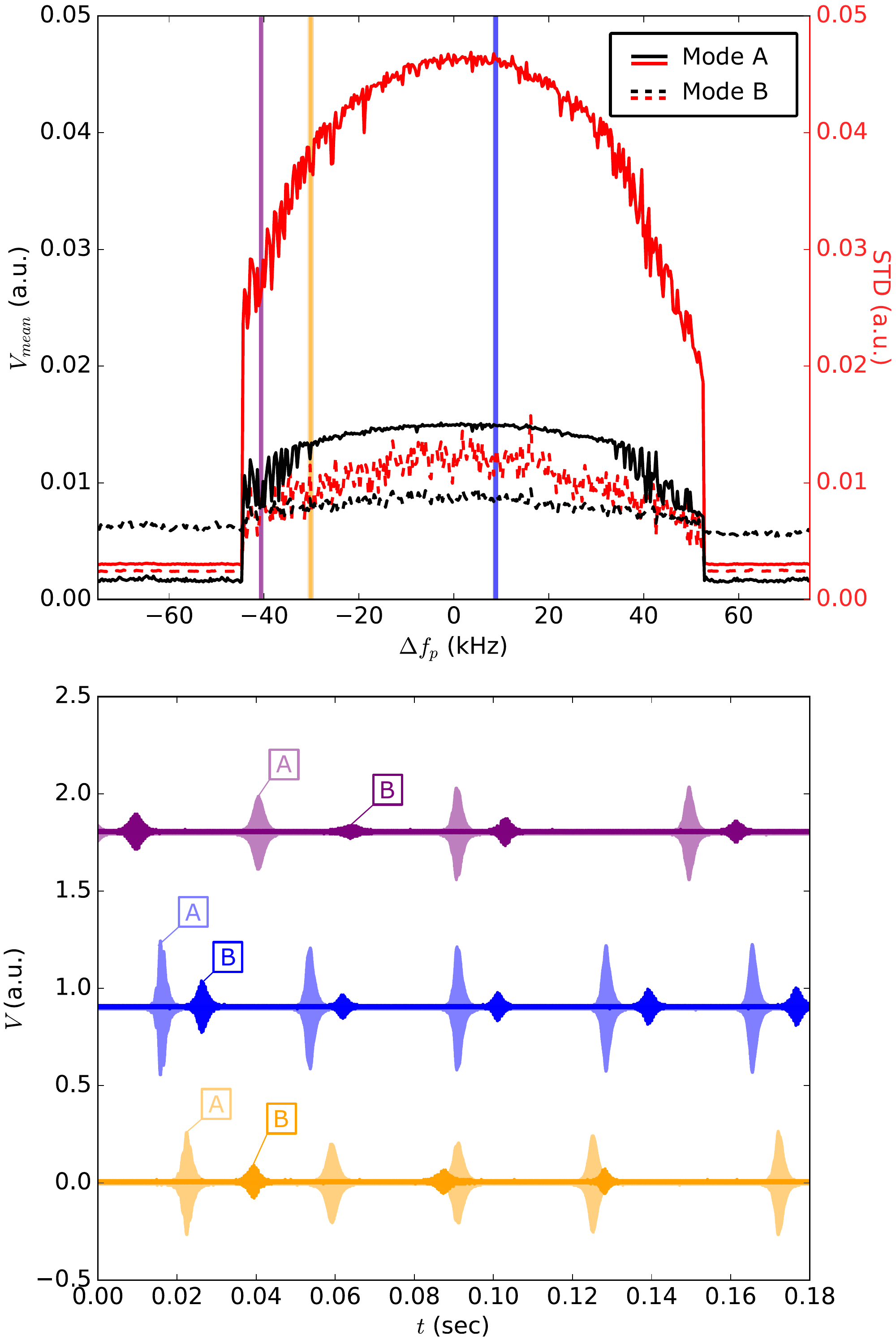}}
\end{minipage}%
\begin{minipage}{.33\linewidth}
\centering
\subfloat[]{\label{}\includegraphics[scale=.28]{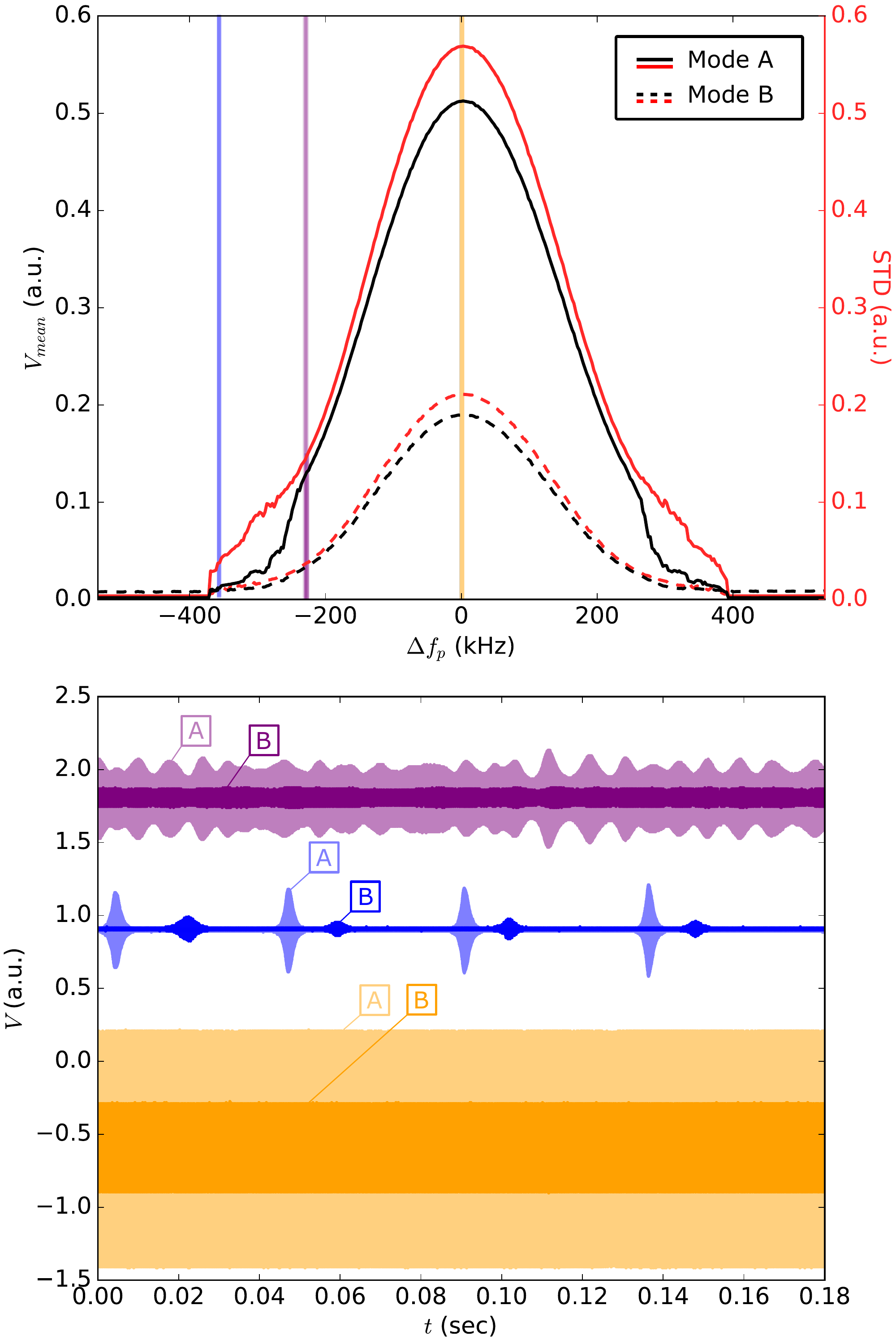}}
\end{minipage}%
\begin{minipage}{.33\linewidth}
\centering
\subfloat[]{\label{}\includegraphics[scale=.28]{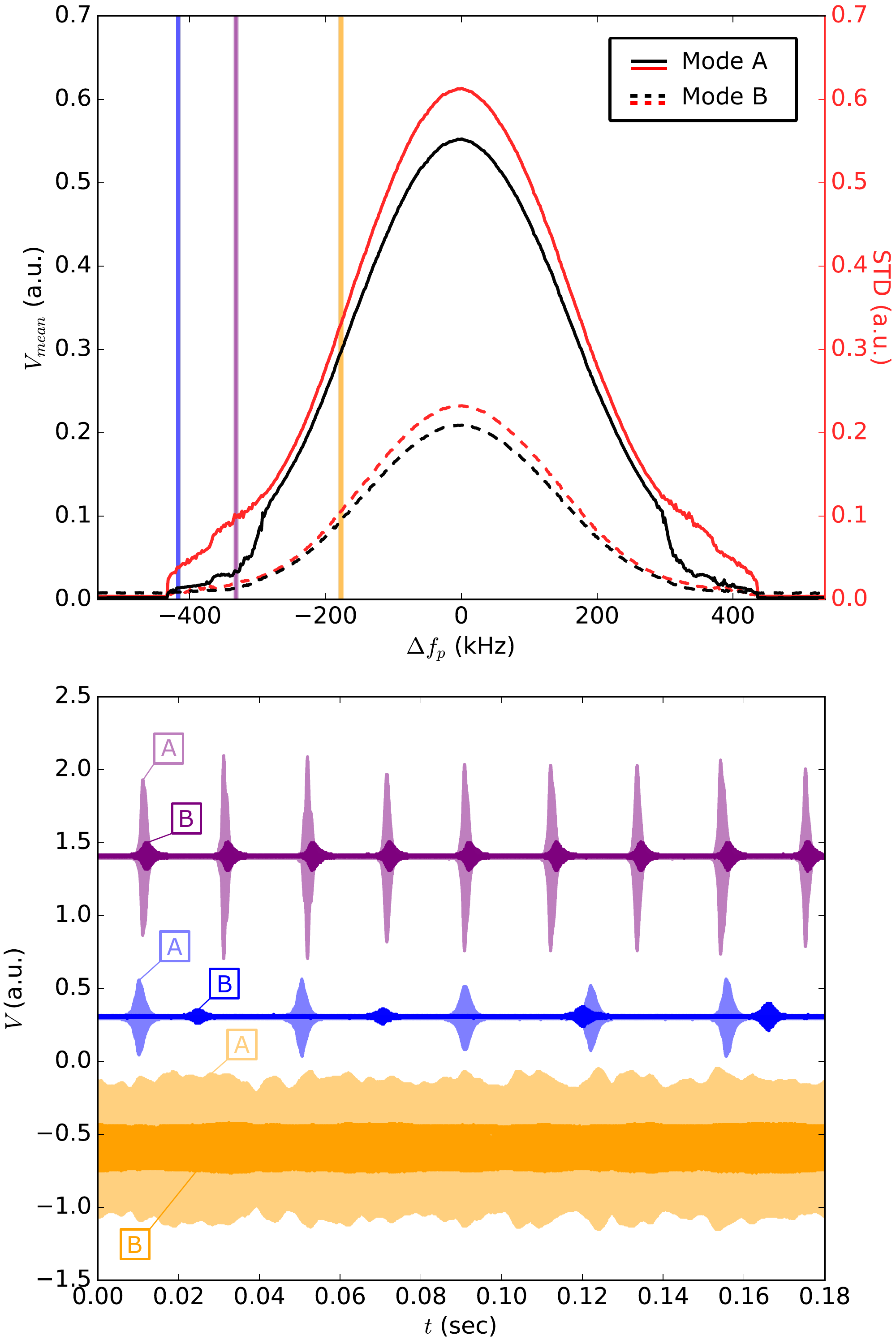}}
\end{minipage}%
\vspace{-3 mm}
\end{figure*}
\begin{figure*}[h]
     \begin{center}
            \includegraphics[width=0.5\textwidth]{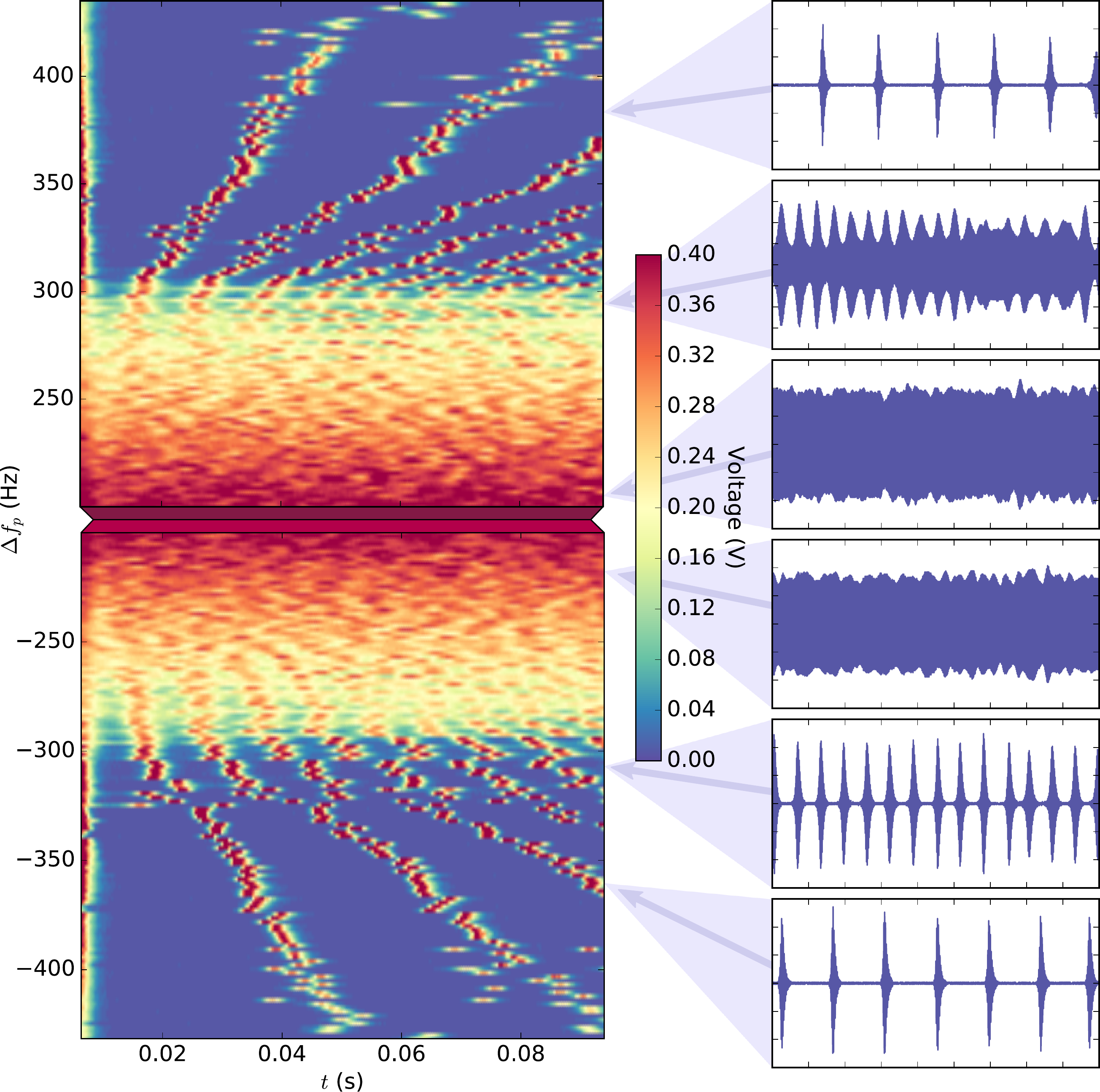}
            \end{center}
\vspace{-5 mm}
    \caption{At incident powers of -41~dBm, -36~dBm and -31~dBm ((a),(b), and (c)), the pump frequency was swept around pump mode 1 ($\Delta f_p$), and solitons were observed at the mixed down mode outputs A and B. The top plots record the standard deviation (red) and mean output voltage (black) of the soliton train (or continuous wave) emanating from the crystal at the signal modes. Depending on the input pump frequency, solitons may be emitted discretely in pulses, or the output may evolve into a continuous wave. The bottom plot (right) presents snapshots of the soliton output produced at -31 dBm incident power (corresponding to (c), output A), for various $\Delta f_p$. In the color density plot (left), traces are aligned to the first pulse to illustrate intersoliton-period dependence upon pump frequency.}
   \label{FIG11}
\end{figure*}
\begin{figure*}[t]
\begin{minipage}{.4\textwidth}
\centering
\subfloat[]{\label{}\includegraphics[scale=0.3]{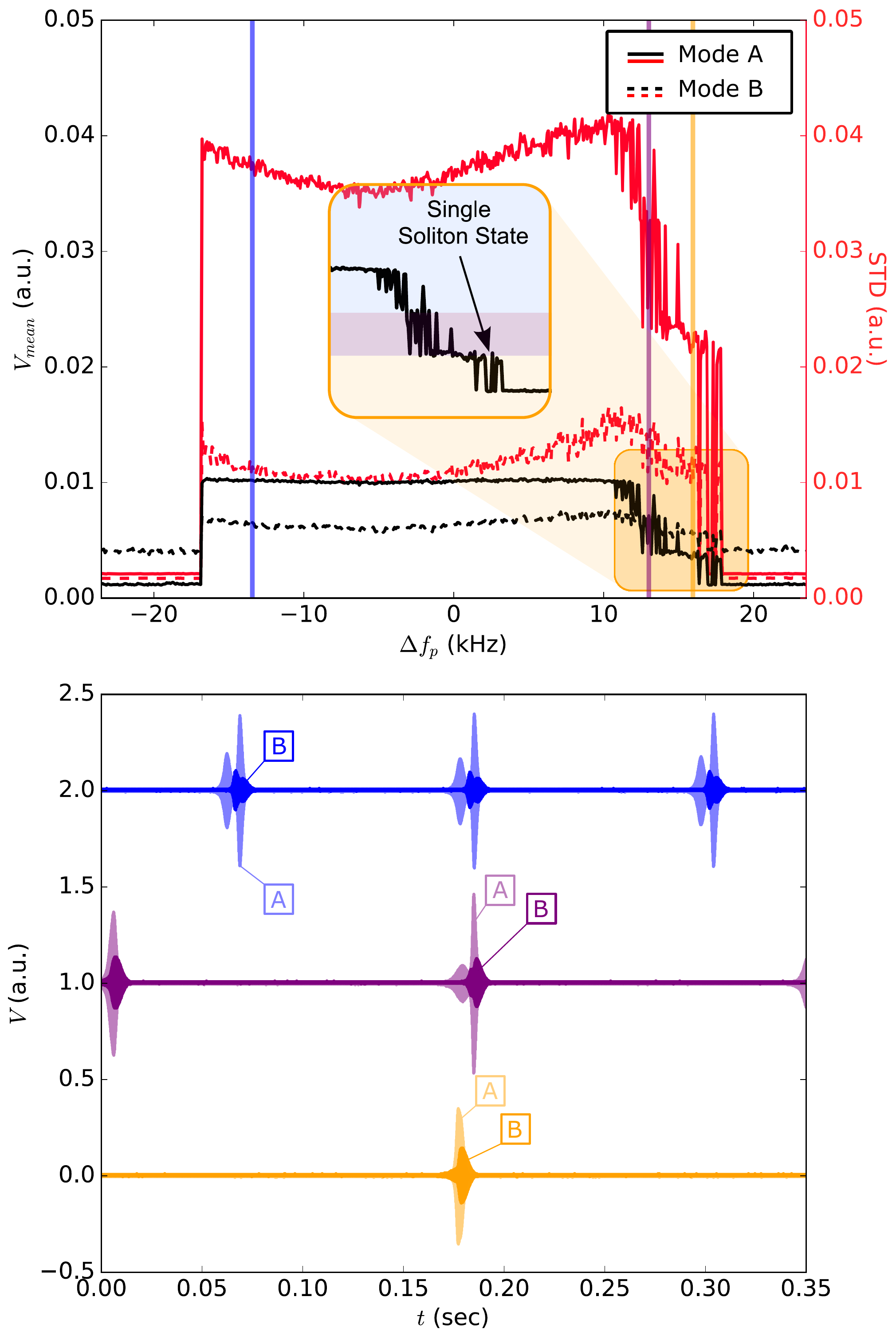}}
\end{minipage}%
\begin{minipage}{.4\textwidth}
\centering
	\subfloat[]{\label{}\includegraphics[scale=0.4]{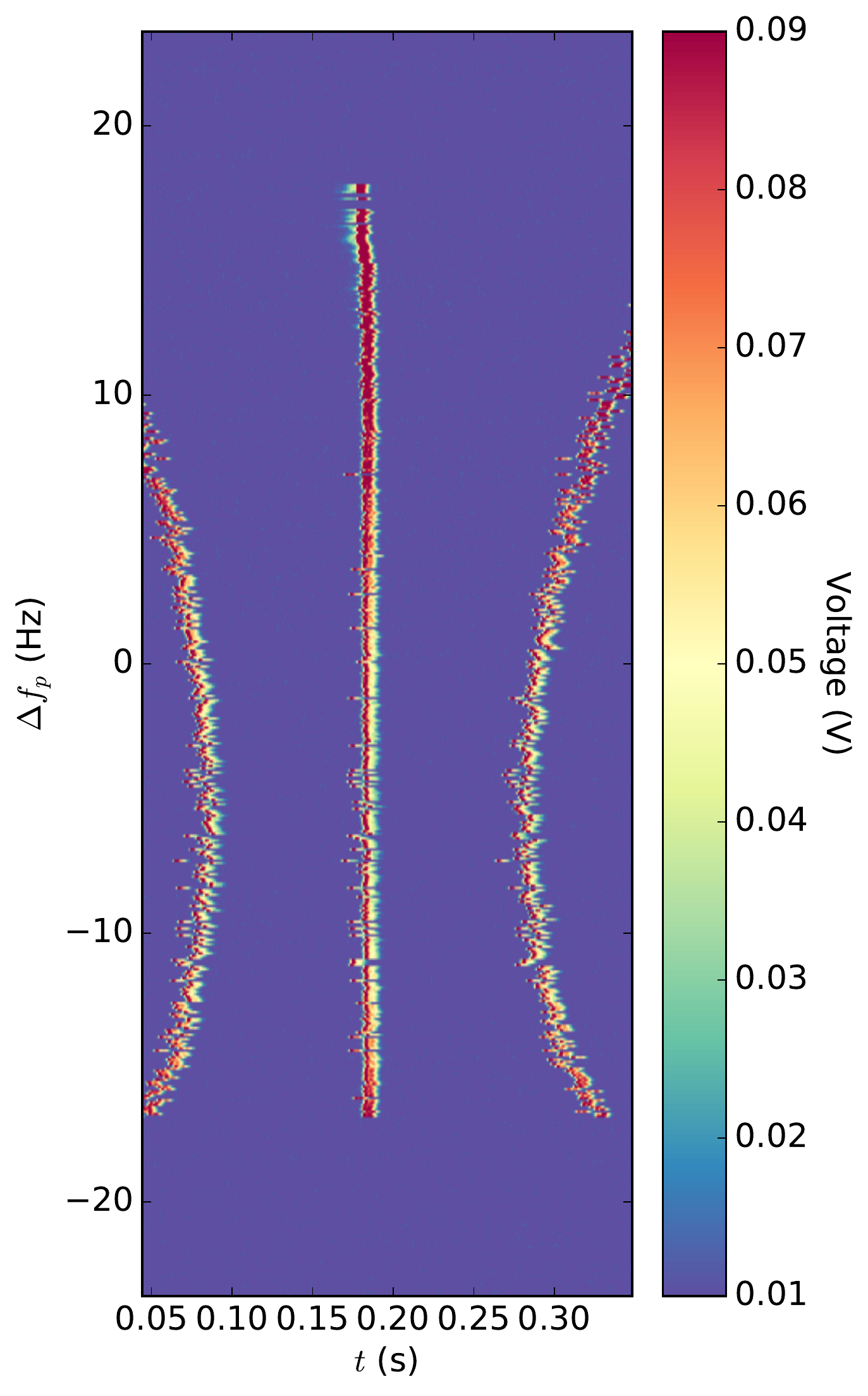}}
\end{minipage}%

\begin{minipage}{.4\textwidth}
\centering
\subfloat[]{\label{}\includegraphics[scale=0.3]{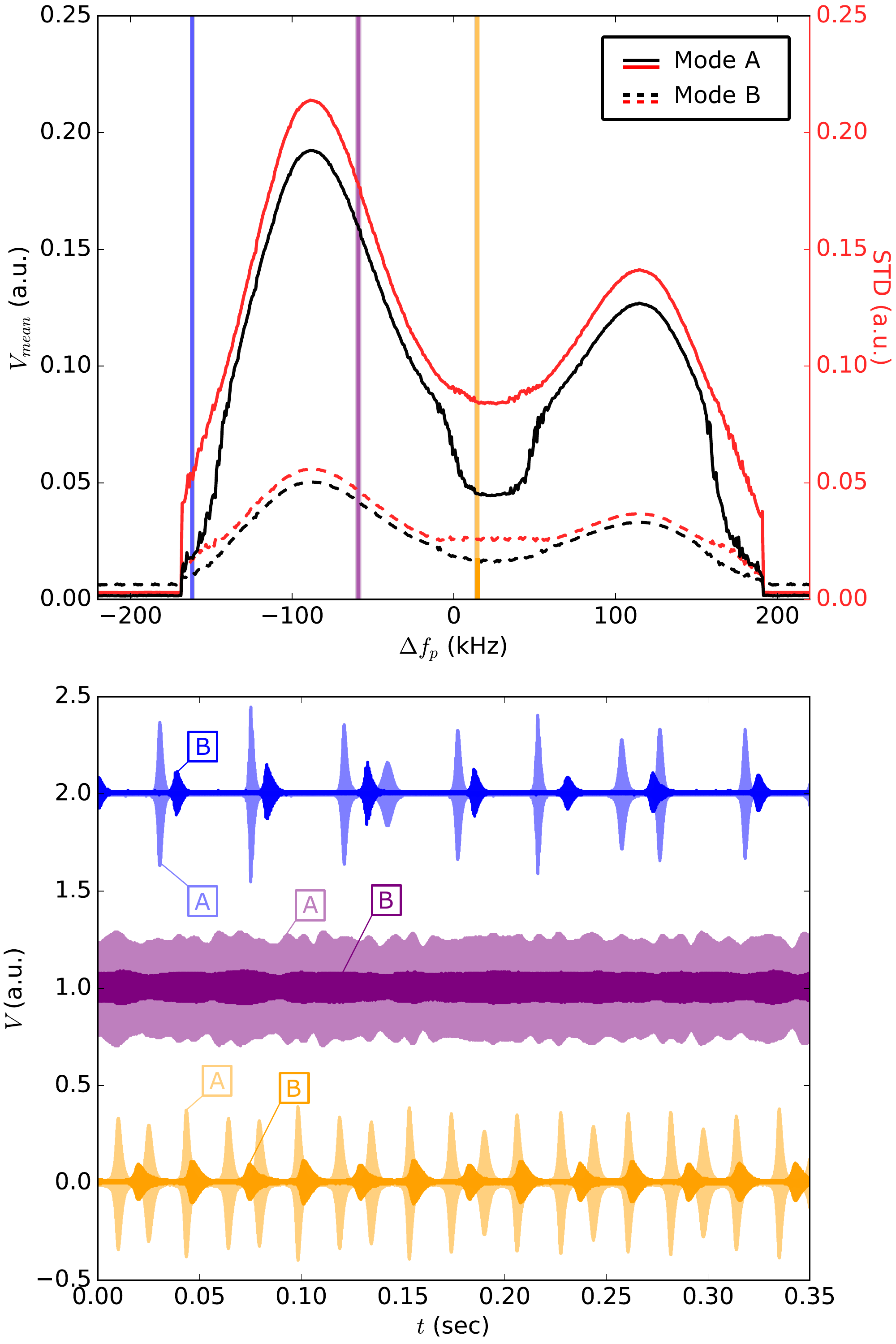}}
\end{minipage}%
\begin{minipage}{.4\textwidth}
\centering
\subfloat[]{\label{}\includegraphics[scale=0.4]{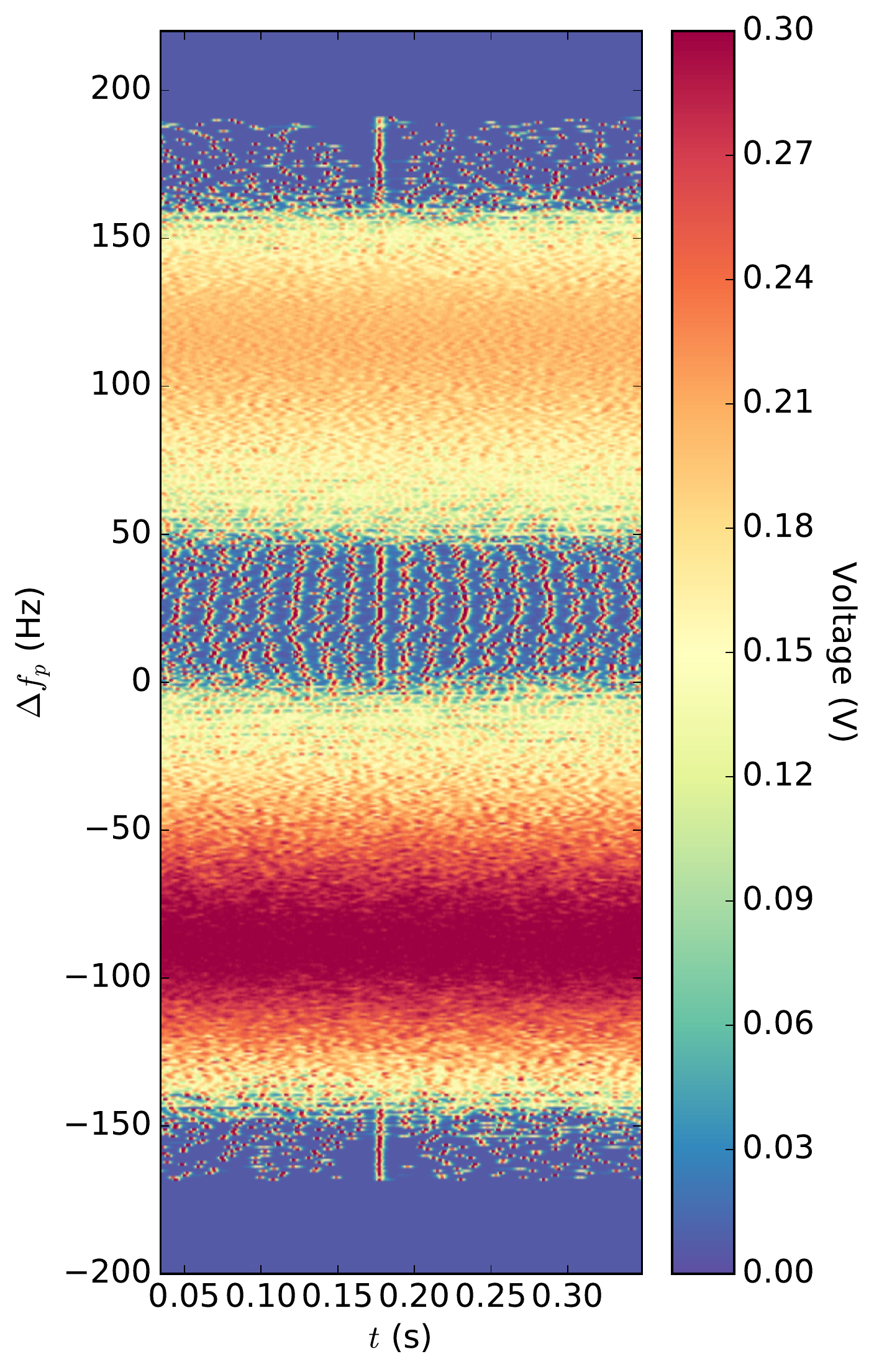}}
\end{minipage}%

\caption{At incident powers of -34.5~dBm ((a) and (b)) and -32~dBm ((c) and (d)), the pump frequency was swept around pump mode 2, and solitons were observed at the mixed down mode outputs A and B. The left plots record the standard deviation (red) and mean output voltage (black) of the soliton train (or continuous wave) emanating from the crystal at the signal modes. Plots (b) and (d) present snapshots of the soliton outputs produced at -34.5~dBm and -32~dBm incident power, from mode A and B respectively,  for various $\Delta f_p$. In the latter, we can observe the pump twice breaching the power level at which discrete soliton emission can be sustained, caused by the doublet shape of the pump mode. Highlighted in (a), the mean output voltage shows clear jumps at certain $\Delta f_p$ steps, which suggests the elimination of one intracavity soliton at each step. The yellow region (corresponding to the yellow time-trace) is therefore representative of the single soliton state.}


\label{FIG12}
\end{figure*}
\clearpage

\bibliography{MaserSolitons}
\bibliographystyle{unsrt}
\end{document}